\documentclass[12pt]{article}

\usepackage{amssymb}
\usepackage{amsmath}
\usepackage{graphicx, color, psfrag, epsfig}
\usepackage{enumitem}
\usepackage[thinc]{esdiff}
\usepackage{natbib}
\usepackage{url} 
\usepackage{verbatim} 
\usepackage{caption}
\usepackage{titlesec}
\newcommand*{\justifyheading}{\raggedright}
\titleformat{\subsection}
  {\normalfont\large\bfseries\justifyheading}{\thesubsection}{1em}{}
\newcommand{\sign}{\text{sgn}}

\global\hyphenpenalty=0

\usepackage{changepage}

\begin{document}

\newcommand{\dblprime}{^{\prime\prime}}

\captionsetup[figure]{labelsep=space}

\changepage{+1.5in}{}{}{}{}{-1.2in}{}{}{}

\begin{titlepage}
\title{\textbf {\Large{Delegation and Participation in Decentralized Governance: An Epistemic View}}}

\author{Jeff Strnad\footnote{Charles A. Beardsley Professor of
Law, Stanford University.  I am grateful for comments from participants at the 2024 Equitable Tech Summit on early versions of some of the results and from Yann Aouidef. Bill Eskridge, Leo Glisic, and Takuma Iwasaki. }}
\date{}

\maketitle  
\thispagestyle{empty}   

\begin{abstract}
We develop and apply epistemic tests to various decentralized governance methods as well as to study the impact of participation. These tests probe the ability to reach a correct outcome when there is one. We find that partial abstention is a strong governance method from an epistemic standpoint compared to alternatives such as various forms of ``transfer delegation" in which voters explicitly transfer some or all of their voting rights to others. We make a stronger case for multi-step transfer delegation than is present in previous work but also demonstrate that transfer delegation has inherent epistemic weaknesses. We show that enhanced direct participation, voters exercising their own voting rights, can have a variety of epistemic impacts, some very negative. We identify governance conditions under which additional direct participation is guaranteed to do no epistemic harm and is likely to increase the probability of making correct decisions. In light of the epistemic challenges of voting-based decentralized governance, we consider the possible supplementary use of prediction markets, auctions, and AI agents to improve outcomes. All these results are significant because epistemic performance matters if entities such as DAOs (decentralized autonomous organizations) wish to compete with organizations that are more centralized.
\end{abstract}

\vspace{4mm}

\begin{center}

\vspace{4mm}

May 6, 2025 Version

\vspace{4mm}


\copyright Jeff Strnad

\end{center}

\changepage{-1.0in}{}{}{}{}{+1.2in}{}{}{}

\newpage
\tableofcontents \thispagestyle{empty}

\end{titlepage}     

\changepage{-1.5in}{}{}{}{}{+1.2in}{}{}{}

\section{Introduction}\label{s1}

Decentralized Autonomous Organizations (``DAOs") operate largely through the execution of code and have no centralized management. When participant decisions are required, DAOs typically utilize majority voting by token holders with one vote per token. In most cases, the voting rights are publicly traded. 

\emph{Decentralization}, defined as the ability for the token project to 
operate in the absence of trusted parties, is a desideratum for DAOs. 
Trustlessness is a major potential benefit because trusted parties may betray 
the trust put in them due to conflicts of interest, bias, or outright 
dishonesty, each of which may imperil the project or diminish its 
performance. 

Despite the potential benefits of trustlessness, there is the possibility that centralized 
approaches will result in more effective project management. If the ensuing 
competitive disadvantage is severe enough, DAOs will not be a viable option.  
In the ensuing debate, a counterargument, appearing, for example, in 
\citet{Buterin_2022_09}, is that in some contexts, decentralized management 
will be superior because of the \emph{wisdom of crowds}. Roughly speaking, 
when decision authority is dispersed, the aggregation of knowledge through 
the collective judgment of the many expressed through voting or otherwise may 
dominate decisions by one or only a handful of managers. 

Motivations for decentralization extend beyond the possibility of exploiting 
the wisdom of crowds. Perhaps most prominent are web3 ideals, well described in \citet{Dixon_2024}. A starting point for these ideals is the observation that during ``web2," the previous stage, a small group of companies gained substantial power and control by leveraging network economies of scale to become the primary parties intermediating the internet. This leverage gave them the power to extract valuable information from large groups of people for free, to censor and exploit creators who depend on internet audiences, to constrain the speech of internet users, and to control or heavily influence what content internet users see. In contrast, a core element of web3 is ``democratizing ownership" with the ultimate goal of placing economic and governance power in the hands of users while retaining the competitive and efficiency advantages of corporate networks. This goal raises exactly the conundrum that is the major focus of this paper: how to create governance that is both decentralized and efficient enough to be competitive with centralized alternatives.

\emph{Participation} has emerged as a major concern for DAOs. For the vast majority of projects, voter participation is very low and in many cases it is clear that a small group of actively engaged parties, including typically, some large holders, predominate in governance.\footnote{\citet[pp. 5-6, 8]{Feichtinger_2023} document very low Nakamoto coefficients, expressing the minimum number of token holders or delegates with more than half the voting rights, for a sample of 21 DAOs, including many of the most prominent ones. 17 of the 21 DAOs have either a delegate or token holder Nakamoto coefficient of less than 10. They also find very low participation rates for token holders. \citet{Liu_2023} finds similar levels of both concentrated control and low participation in a sample of 50 DAOs, and \citet{Fritsch_2024} identify similar patterns for Uniswap and Compound in a more detailed examination of 3 DAOs.} There are many sound reasons not to vote. Consider three of them. First, small holders have little economic incentive to become informed. As a result, if there are only small holders, there is likely to be a serious collective action problem. Relatively few voters become well informed, and DAO functioning suffers. Second, and related, small holders face the ``paradox of voting." Voting is costly, but with a small stake, there is only a very small chance that the stakeholder's vote will be decisive. Third, some DAO token owners will be portfolio investors, holding a large number of different positions for the purpose of diversification. It will be difficult and costly for them to participate in governance of all or even more than a few of the projects. In addition, an important motivation for diversification is to avoid relying on superior outcomes for particular projects and the consequent need to study and choose among them. 

Prominent academics and industry participants see explicit delegation of voting rights to others as a possible way both to address low participation and to direct voting power to more informed and competent voters.\footnote{See, e.g., \citet{Hall_2024}(discussing the idea, some of the associated literature, and providing a cogent empirical analysis of the operation of liquid democracy in DAO governance, including the impact on participation).}  In addition, there has been substantial interest in the nature of delegation itself and in possible innovations such as the use of liquid democracy, which permits unlimited delegation and re-delegation of votes.

In the rest of the paper, we focus on decentralization, the wisdom of crowds, 
participation, and delegation. We claim that the relationship between them is 
more complex than appears from the existing literature. Decentralization does 
not necessarily advance the wisdom of crowds. Many methods of delegation are 
problematic. Increased participation is not necessarily desirable.\footnote{Some prominent parties active in the DAO space have expressed skepticism about promoting participation in an undifferentiated way. For example, \citet{Spannocchi_2025}, in thoughtful article, describes various DAO participants and discusses in detail the pluses and minuses of their participation in governance. The results here are very consistent with that kind of nuanced viewpoint.} Methods that include partial abstention may be superior to common approaches that rely on explicit delegation of voting rights to others.

The analytic approach underlying these claims is based on two \emph{epistemic tests} and is closely related to the wisdom of crowds. These tests consider a choice between two alternatives, one of which is correct, and assess how well various governance approaches aggregate information from a diversity of voters to maximize the probability of making a correct collective decision. The first test assumes that voter judgments are independent of each other. The second test permits any pattern of dependencies. Regardless of the test, we make very strong assumptions about each voter's ability to assess the quality of their information. In particular, voters can translate any collection of information into a precise estimate of the probability of reaching the correct collective decision based on that information.

This framework is extremely limited because of the associated strong assumptions, including assuming that there is a correct answer and that voters are perfect judges of the quality of their information. In effect, the tests we apply make it particularly easy for governance approaches to be successful. If a governance approach performs poorly in these simple situations, it is unlikely to be a good or useful approach. The tests in fact reveal substantial weaknesses in some prominent approaches. In addition, the tests create considerable insights about such approaches that transcend the limitations inherent in the framework.  

The next section describes the two tests and the underlying epistemic framework and concludes with a terminology subsection. The third section discusses methods that rely on partial abstention, the ability of voters to exercise only a portion of their voting rights. The fourth sections studies participation by voting directly, especially the epistemic impact of adding new voters. The fifth section focuses on explicit delegation of voting rights to others. The sixth section examines some non-voting supplements to 
token-based voting arrangements that may significantly improve governance. A final section concludes.

The results and analysis in the paper apply to decentralized governance in general, whether or not implemented using DAOs. Nonetheless, we pay special attention to DAOs both because they are important applications of decentralized governance and because a focus on DAOs often facilitates the exposition of more general ideas.

\section{Analytic Framework}

\subsection{An Epistemic Approach}\label{epistemic}

One way to formalize the wisdom of crowds phenomenon is through ``jury theorems." A good conceptual starting point is the earliest result due to \citet{Condorcet_1785}, often called \emph{Condorcet's Jury Theorem}. This result assumes a dichotomous choice by majority vote, which is a \emph{determinable decision} in the sense that one of the two alternatives is the right one and the other is wrong. 

Condorcet's Jury Theorem consists of the following. Assume that voters have 
equal probabilities, $p = 0.5 + \epsilon$ for some $\epsilon > 0$, no matter how small, of making the right decision and that their judgments are statistically independent. Using the terminology of \citet{Dietrich_2023}, two conclusions follow: 

\begin{enumerate} [label=\arabic*)]
\item \emph{Increasing Reliability}: The probability that the majority-vote 
    collective decision will be correct is increasing in the number of 
    voters. 
\item \emph{Asymptotic Infallibility}: As the number of voters increases  
    to infinity, the probability that the collective decision will be 
    correct converges to 1. 
\end{enumerate}

\noindent Thus, majority voting by a sufficiently large number of independently informed people each of whom is only a little better than a coin flip at coming up with the correct alternative will dominate any finite group of decisionmakers unless at least one decisionmaker in that group has \emph{substantive omniscience}, that is, $p=1$. 

Although Condorcet's Jury Theorem is a cogent conceptual illustration of the wisdom of crowds, it is limited in scope due to the Theorem's restrictive assumptions: All voters have the same competence, the correctness of judgments are statistically independent, and the asymptotic result depends on having a very large number of individuals. Both epistemic tests we develop relax the assumption of equal competencies and one of them relaxes the assumption of statistical independence. Furthermore, most of the results we derive do not depend on the number of voters.

The epistemic framework that applies to both tests and much of the ensuing analysis relies on several other major assumptions. We list and discuss them in the rest of this subsection. A first assumption we already have encountered:\\

\textbf{Assumption 1 (Determinable Decisions):} \emph{Decisions are determinable. There is a single correct answer.}\\

This assumption is a substantial limitation. \citet[p. 27]{Dietrich_2023} note that the idea that ``alternatives are factually correct or incorrect" is a ``controversial philosophical premise." They discuss some of the literature surrounding the question of whether ``correctness facts" and thus determinable decisions exist, especially in domains that involve political or moral decisions or a mixture of empirical and normative elements. 

Nonetheless, there are two reasons to proceed with this assumption. First, we would like DAOs to perform well when the decisions they make \emph{are} determinate. Certain technical questions, for example, plausibly have a correct answer, albeit subject to current uncertainty. Second, we will show in subsection \ref{contestable} that this assumption may be appropriate in the case of what we call ``economic DAOs." For these DAOs, the market token price can embody many normative, political, or moral elements such as the value of participation for which there is no correct answer at the level of reasoning alone. In the case of economic DAOs, the right answer arguably is the one that maximizes the market price.

In their overview of jury theorems, \citet[p. 35]{Dietrich_2023} point out that ``conventional jury models implicitly assume that individuals share the same objective of a correct collective decision." That is our second assumption:\\

\textbf{Assumption 2 (Shared Epistemic Objective):} \emph{Voters share the same epistemic objective: reaching correct collective decisions.}\\

This assumption makes it much easier to define an effective governance mechanism in at least two significant ways. First, the assumption rules out malicious actors and actions. It is unrealistic to ignore potential malicious behavior and attacks, but not having to address these potentialities makes the governance design problem much simpler.

Second, the assumption means that we do not have to worry about the possibility that some voters may choose to assert their own beliefs about correctness without giving due weight to others in a way that will maximize the probability of a collective decisions being correct. Instead, voters will try to facilitate inclusion of contrary views if that will increase the likelihood of a correct collective decision.

A different situation occurs when voters have \emph{instrumental objectives} instead of a shared epistemic objective. Asserting one's own beliefs is one example of an instrumental objective. More generally, when there is no correct collective decision, the social choice process is at least in part engaged in preference aggregation rather than information aggregation. In that situation, voters will focus on clarifying and promoting their preferences. For example, \citet{Bloembergen_2019} model voters who are not sure of their preference in a binary choice and delegate to others who are likely to discern and vote for their preference. A few of the results in later parts of this paper will apply both when there is a shared epistemic objective and when voters implement instrumental objectives. We will note when that is the case. In many other instances, Assumption 2, that there is shared epistemic objective, will be a necessary condition for the results to hold.

A third major assumption concerns the voters that comprise the governing population along with their information and beliefs:\\

\textbf{Assumption 3 (Cognizable Probabilities):} \emph{Voters can evaluate their total information set or any subset thereof and compute an accurate probability that they can choose the correct decision alternative based on that set or subset of information.}\\

This assumption is very strong and has major implications. First, there is no reason to worry about voters being more likely wrong than right when they vote. Under the assumption, a voter who is thinking of voting for A over B based on particular information knows the probability of that inclination being correct. If the probability is less than $0.50$, then they will vote for B given that they share the epistemic objective of promoting the correct collective decision. Each vote based on each information set will have a probability $\geq 0.5$ of being correct. 

This implication of the assumption is important because the danger that voters have a degree of ignorance that makes them less likely than a coin flip to be correct is a major concern for voting systems and makes it much harder to design an effective governance mechanism. For example, it is an element of the critique of majority voting in \citet{Brennan_2016} based on the associated risk that majority voting will lead to defective outcomes. In the context of Condorcet's Jury Theorem, if the common probability of each voter being correct is less than $0.5$, then adding voters will increase the probability of making the wrong decision and that probability will converge to one as the number of voters increases without bound. 

Second, this assumption means that when two voters receive the same information, they will agree on the implications of that information for the probability of voting for the correct decision. This property makes it much easier to address the situation in which voter judgments are not statistically independent. A further assumption also facilitates addressing dependencies:\\

\textbf{Assumption 4 (Information Decomposition):} \emph{The totality of information received by voters can be decomposed into a ``canonical list" with two properties. First, each item of information on the list is statistically independent from all other items on the list. Second, each voter's information set is a sum of items on the list.}\\

This assumption allows us to address the situation of dependencies between voter judgments by focusing instead on a series of independent signals, the items on the canonical list. This approach greatly simplifies the analysis, provides great clarity with respect to some governance approaches, and suggests ways of addressing dependencies in the field.\footnote{The independent signals approach here is inspired in part by \citet{Dietrich_2024}, who developed a similar approach to address the epistemic impact of deliberation in a one-person-one-vote majority voting context. Deliberation creates dependencies when individuals share sources. \citet{Dietrich_2024} show that the overall impact of deliberation is not clear. Consider two simple examples, one negative and the other positive. If a particular signal, $x$, spreads easily through deliberation and other signals do not, the epistemic quality of collective decisions can suffer. If the result is that most or all voters receive $x$, then it will tend to drown out the other signals. The outcome can be a dependency pattern such as the one described in Example 2.1 on page \pageref{dependency_example} infra, which, as the example indicates, can have a very negative epistemic impact. In contrast, if deliberation causes all voters to receive all the signals on the canonical list, then every voter can make the best possible collective decision unilaterally, and, given the shared epistemic objective, any vote will unanimously favor that decision.}

These four assumptions are very strong both individually and collectively. Separately and together, they make it much more likely that governance approaches will be effective. That thumb on the scale is a major defect if we base an argument for the effectiveness of any approach on an analysis that relies on the assumptions. However, what we find is that many major governance approaches have substantial weaknesses even in the face of these helpful assumptions. 

Finally, there is a more general purpose for using these restrictive assumptions as well as for focusing on two epistemic tests rather than a more general model. The assumptions and the tests create a simple enough context to make the underpinnings of various results very clear but rich enough to have obvious generality. At some points, we consider extensions that broaden the context. Throughout, we use examples in an attempt to create clear understanding and intuition.

\subsection{A Base Model and Two Epistemic Tests}

\subsubsection{Base Model}

Define $V$ to be the \emph{voter profile}, a list of $N$ \emph{voters} indexed by $i$ of all persons who have voting rights through token ownership or delegation along with the number of voting rights, $t_i$, that each such person $i$ holds. Note that delegated voting rights as of any point in time are counted as being held by the delegate and not the delegator. ``Persons'' include entities as well as individuals. We say $i \in V$ if individual or entity $i$ is on the list. 

These persons vote on a dichotomous choice between policies A and B, one of which is the correct collective choice under the determinate decision assumption.  Voter $i$ has a probability $p_i$ of making the correct decision. We say that this probability represents that person's \emph{epistemic competence}. Under the cognizable probability assumption each voter $i$ knows the probability $p_i$.

A person $i$ may choose to vote only $\bar{t_i} \in [0, t_i)$ of their voting rights. In that case, we say that person $i$ \emph{abstained} from voting on $t_i - \bar{t_i}$ of their voting rights. Taking into account abstentions, we form $\Bar{V}$, the \emph{vote profile} consisting of the same list of persons in $V$ along with the actual number of voting rights, $\bar{t_i}$, that person $i$ has exercised. We say $i \in \Bar{V}$ if $\bar{t_i} > 0$ for person $i$. We do not distinguish between abstention by passivity, not voting at all, and abstentions in which person $i$ does vote but specifies ``abstain" as a ballot choice.

In most situations, we will assume that we are at the point of decision so that voters cannot delegate their voting rights to others, no persons may be added to the voter profile, no further deliberation will occur, and voters cannot increase their probability of being correct by learning new information. But we will consider approaches such as liquid democracy where further delegation is possible, and we will consider the impact of expanding the voter profile by adding new voters or delegates.

\subsubsection{Independent Competencies Test}

The independent competencies test relies on the following assumption:\\

\textbf{Assumption 5A (Independent Competencies):} \emph{Voter judgments are statistically independent.}\\

This assumption gives us a powerful tool to assess governance approaches. As shown by 
\citet{Nitzan_1982}, if voters have independent competencies, then optimal voting weights are:
\begin{equation} \label{optimal_weights} 
w_i = ln\left(\frac{p_i}{1-p_i}\right).
\end{equation}
Here ``optimal" means creating the highest possible probability of a correct 
collective decision. We refer to this optimality result as the \emph{Optimal 
Weighting Theorem}.

Let $v_i$ indicate person i's vote where: 
\begin{equation} \nonumber
    v_i=
        \begin{cases}
                +1 & \text{if voter i votes for A}\ \\
                ~~0 & \text{if voter i abstains}\ \\
                -1 & \text{if voter i votes for B}\ 
        \end{cases}
 \end{equation}
\noindent Then the collective decision, $d$, with the highest probability of being correct is:
\begin{equation} \nonumber
    d=
        \begin{cases}
                \text{A if}\ \sign\left(\sum_{i=1}^{N} w_i v_i\right) >0 \\
                x \in_R \{A,B\}~ \text{if}~ \sum_{i=1}^{N} w_i v_i = 0 \\
                \text{B if}\ \sign\left(\sum_{i=1}^{N} w_i v_i\right) <0 
        \end{cases}
 \end{equation}
\noindent where $\in_R$ indicates taking a random draw uniformly from a 
set.\footnote{In other words, we resolve a tie vote through a coin flip.} A useful property 
of the weights in this setting of independence that we are assuming is that 
the weights depend only on each voter's epistemic competence and not 
on the characteristics of any other voter. 

To explore the meaning of the weights, consider the following table. The first 
column contains various weights, and the second column indicates the probabilities 
corresponding to the weights. In a situation of \emph{ignorance}, a voter  can do no better than a coin flip in deciding between A and B. 
The probability of being correct is $0.5$, and the optimal weight is 
zero.\footnote{The same probability and weight apply if the voter is 
informed but ends up being equally balanced between A and B.}\\  

\begin{table}[htbp]
  \centering
  \caption{Weights for Probabilities}
    \begin{tabular}{ccccc}
    \multicolumn{1}{l}{weight} & \multicolumn{1}{l}{probability} & \multicolumn{1}{l}{3 voters} & \multicolumn{1}{l}{5 voters} & \multicolumn{1}{l}{7 voters} \\
          &       &       &       &  \\
    -2    & 0.1192 & 0.0392 & 0.0141 & 0.0052 \\
    -1    & 0.2689 & 0.1781 & 0.1245 & 0.0894 \\
    -0.1  & 0.4750 & 0.4626 & 0.4532 & 0.4455 \\
    0     & 0.5000 & 0.5000 & 0.5000 & 0.5000 \\
    0.1   & 0.5250 & 0.5374 & 0.5468 & 0.5545 \\
    1     & 0.7311 & 0.8219 & 0.8755 & 0.9106 \\
    2     & 0.8808 & 0.9608 & 0.9859 & 0.9948 \\
    4     & 0.9820 & 0.9990 & 0.9999 & 1.0000 \\
    5     & 0.9933 & 0.9999 & 1.0000 & 1.0000 \\
    10    & 1.0000 & 1.0000 & 1.0000 & 1.0000 \\
    \end{tabular}%
  \label{Table_1}%
\end{table}

If the voter is more likely to be wrong than correct, the voter's weight is negative. In that case, the voter's choices still contain valuable information. The negative sign simply shifts the valence of the choices in the tabulation. Examination of the table reveals that probabilities equally above and below $0.5$ result in weights with the same absolute value, but opposite signs. The corresponding votes contain equally useful information.  

The final three columns in the table state the probabilities of a correct 
collective decision with majority voting for 3, 5, and 7 voters with independent competencies, all with the 
same epistemic competence indicated in the second column. These columns 
demonstrate the increasing reliability, or in the case of probabilities below 
$0.5$, the decreasing reliability, that results from increasing the number of 
voters, as in Condorcet's Jury Theorem. 

\subsubsection{Dependent Competencies Test}

Assuming independent competencies simplifies the analysis because it allows us to apply the Optimal Weighting Theorem directly. But independence of competencies is a very strong assumption. We would expect voters to share some information sources, creating dependencies in competencies. In this subsection, we drop Assumption 5A of independent competencies in favor of a less restrictive alternative:\\

\textbf{Assumption 5B (Dependent Competencies):} \emph{Voter judgments may be statistically dependent.}\\

Dependent competencies can greatly affect the analysis as indicated by the following example:\\

\textbf{Example 2.1.} \label{dependency_example} There are 3000 voters with weight $w_i=0.1$ ($p_i \approx 0.525$), and one voter with weight $w_i=1$ ($p \approx 0.731$). Weighted voting using decision rule $d$ determines the collective decision. Assuming independent competencies for the 3000 voters, the probability that the collective decision will be correct is approximately $0.997$ if only those voters participate, Condorcet's Jury Theorem in action. Adding an independently competent voter with weight $1$ increases this probability by only $.000028$, leaving $0.997$ as the rounded result.\footnote{Under weighted voting, with probability $0.731$ the more competent voter will correct results in which between 1490 and 1500 of the other voters choose the right alternative but with probability $0.269$ the more competent voter will reverse a correct collective outcome when between 1500 and 1510 of the other voters choose the right alternative. The calculation of these ranges and effects follows from application of the cumulative binomial distribution.} Suppose, however, that the judgments of the 3000 voters are perfectly correlated. This dependence might result from relying on a single source, such as consulting a single prominent internet or Discord post in the absence of having any expertise. Now the 3000 voters are equivalent to a single voter in value added and deserve a cumulative weight of only $0.1$. The more competent voter has ten times this weight and will unilaterally determine the outcome under decision rule $d$. The probability of a correct collective outcome plunges from $0.997$ to $0.731$, the epistemic competence of the more competent voter acting alone. If we fail to allow for dependencies by continuing to weight the set of less competent voters as if they had independent competencies, the outcome is even worse. We will incorrectly weight the 3000 voters three hundred times more than the single more competent voter. Because the 3000 voters act as if they were a single voter due to the perfect correlation, the probability of making a correct collective choice plunges even further to $0.525$. This effect of failing to adjust the weights to take into account dependencies can occur with as little as eleven less competent but perfectly correlated voters combined with the single more competent voter. $\square$\\

The example not only indicates the need to adjust the weighting method to allow for dependencies, but also is suggestive of how to do so. With independent competencies, each vote is an independent signal of which collective alternative is the correct one. The Optimal Weighting Theorem tells us how to weight the various signals based on each signal's probability of indicating the correct collective outcome by itself. In Example 2.1 under the dependence assumption, there are in effect only two signals, one that should be weighted $1$ and the other $0.1$. One way to reach that result would be to create adjusted weights by dividing the optimal weights of each of the 3000 perfectly correlated voters that would obtain under independent competencies by 3000. We state this idea, which follows directly from the Optimal Weighting Theorem by simply considering voters to be information signals in that Theorem, formally:\\

\textbf{Proposition 2.2.} \emph{Given Assumption 4 (Information Decomposition), the following method will maximize the probability of making a correct collective decision:}\\

\noindent \emph{(i) For each information item $i$ on the canonical list of independent and exhaustive items, compute the probability, $p_i$, that the correct collective decision will follow solely from the information from that item.}\\

\noindent \emph{(ii) Compute a weight, $w_i$, for each such item in the same manner as under the Optimal Weighting Theorem, that is:} 
$$w_i = ln\left(\frac{p_i}{1-p_i}\right).$$\\

\noindent \emph{(iii) For each information item, determine which collective decision the information for that item indicates is the correct one, tabulating these indications as ``votes" for one alternative or the other.}\\

\noindent \emph{(iv) Combine the weighted votes to arrive at collective decision $d$ in the manner indicated by the Optimal Weighting Theorem.}\\

We have an approach for both the case of independent competencies and the case of dependent competencies of maximizing the probability of making the correct collective decision. A social planner who had all the relevant information about voters or signals could apply those approaches directly.  But in the context of DAO governance, we face a very different problem. Voters are attempting to arrive at the correct collective decision in a decentralized way. The rest of the paper concerns the ability of various governance methods that are or have been considered to be strong candidates for governing DAOs to actualize this goal. We have created two epistemic tests for those methods, one for independent voter competencies and one for the case where those competencies are statistically dependent. The dependent case is much more challenging because of the need to decompose information sets into independent components.\footnote{The limitation to ``candidate" governance methods is analytically significant. Assumptions 2-4 allow us to reach interesting conclusions about these methods, the point of the paper, but the assumptions are so strong that it is possible to create simple decentralized mechanisms that would replicate the performance of a social planner with complete information about voters or signals. For instance, we could create code that implemented decision rule $d$, and then embed that code in a smart contract. Participating voters would report their independent competencies or their signals and the competency associated with each signal to the smart contract along with the vote associated with each element. The smart contract would use the competencies and the set of voter or signal votes to compute the outcome. By Assumption 2, voters would be motivated to report and to report honestly. Assumptions 3 and 4 guarantee voter knowledge of all the relevant competencies and that there is a canonical list of signals that can for the basis of a decision.\label{candidates_only}}

In the next three sections, we discuss abstention, participation, and delegation in that order, an order which allows us to create a sequence of discussions that build upon each other. Throughout, as a default, we assume that voters have flexibility with respect to abstaining, voting, and delegating:\\

\textbf{Assumption 6 (Voter Flexibility):} \emph{Voters may choose to vote  or abstain on any portion of their voting rights. If explicit delegation of voting rights to others is permitted, voters may delegate any portion of such voting rights in any mixture to as many delegates as desired.}\\

Thus, we do not limit voters by an all-or-none choice between abstaining, voting, and explicit delegation of token rights to others. Voters can choose any mixture. In addition, if explicit delegation is permitted, voters are not limited to delegating to only one delegate but can choose any combination of delegates and amounts allocated to each delegate.

Much of the literature considers methods that are not consistent with this assumption, and, at some points in the paper, we examine the exclusive use of one method. We note when that is the case, but, otherwise, we allow ourselves the liberty of the assumption. 

In all three sections we impose decentralization as a constraint. The autonomous actions of the voters solely determine the outcome. Our approach is to apply the two epistemic tests, mining the results for more general insights. Before proceeding, we clarify our terminology.

\subsection{Coherent Terminology}\label{terminology}

As mentioned in the introduction, one idea is to advocate for more 
intensive use of ``delegation" to promote ``participation." This idea is 
curious in light of the fact that abstention itself is a form of delegation. 
By abstaining, one delegates the decision to the remaining voters in whatever 
voting power configuration exists.\footnote{\citet{Feddersen_1996} point out this equivalence in a paper that shows that abstention can be an optimal method of delegation when the other voters have superior information. It is likely that others noted the equivalence prior to 1996, perhaps all the way back into antiquity, but the author has not attempted to track down earlier instances. Current scholars are aware of the equivalence. See, e.g., \citet{Mooers_2024} who compare liquid delegation to majority voting and to majority voting with the option to abstain.} Abstention is not an alternative to delegation but just selection of a different set of delegates. It also is functionally equivalent because the same effect as abstaining can be created by a particular pattern of formal delegation of one's voting rights to others. Exercising one's voting rights directly can be considered a form of delegation, namely self-delegation. It is functionally equivalent delegating to a slate that includes oneself with an allocation of zero to all the other delegates. It seems that every action and inaction can be considered to be ``delegation."

``Participation" suffers from its own considerable ambiguities. It can mean delegating voting rights to others or voting oneself. It also is not clear why it should not also encompass abstention, especially deliberate and considered abstention of the kind discussed in \citet{Feddersen_1996} that may increase the probability of a correct collective decision. Even abstention based on rational passivity arguably is ``participation."  It may be that every action and inaction is ``participation" as well as ``delegation."

In the face of these ambiguities, what terminology shall we use? Two elements are important: the mechanism that embodies the choice process, and epistemic distinctions. Classic delegation involves the explicit transfer of voting rights to others, which we call ``transfer delegation." At the epistemic level unless a transfer exactly replicates the existing voting rights proportions held by the other voters, the transfer will shift the relative voting rights among these voters, an ``external relative weight intervention." Knowing whether a particular external relative weight intervention will improve the quality of the collective choice process requires the delegator to know something about the characteristics of the other voters, for instance, their optimal voting weights. In section \ref{delegation}, we describe this knowledge as an aspect of \emph{delegative competence} as distinguished from epistemic competence, the probability that the voter acting alone will reach the correct collective decision.

The situation is very different with respect to a voter abstaining or exercising voting rights directly by voting. These actions do not affect the relative weights of the other parties. But exercising voting rights directly increases the relative weight of the voter compared to all other voters, which can be thought of as ``voter relative weight assertion."  

Now we specify terminology conventions for the rest of the paper. The next section examines partial abstention. Voter $i$ has $t_i$ voting rights, abstains on $a_i$ of those rights, and exercises $\bar{t_i} = t_i - a_i$ of the voting rights directly by voting where $\bar{t_i} \in [0, t_i]$. As we have seen, this can be considered a kind of delegation, and we title the next section ``Delegation by Partial Abstention." Whenever we use the term ``partial abstention," the delegative aspect is implicit, and voter relative weight assertion occurs to the extent voting rights are exercised directly.

We use the term ``direct participation" to describe exercising voting rights directly by voting. Direct participation is the subject of section \ref{participation}. Other types of participation such as transfer delegation or deliberate and considered abstention are therefore instances of ``indirect participation."

In section \ref{delegation} we study transfer delegation. Although it is possible to create a set of transfers that leaves the relative weights of other voters undisturbed, the focus in section \ref{delegation} is the more general situation in which transfers shift those relative weights. In that general situation, transfer delegation includes an element of external relative weight intervention. 

\section{Delegation by Partial Abstention}\label{abstention}

As discussed in the previous subsection, abstention is equivalent to an actual delegation of one's voting rights proportionally to all other voters. We embody this result in a more general form that includes 
partial abstention as a possibility for all voters:\\ 

 \textbf{Observation 3.1.} \emph{For voter i, abstaining from voting a 
 number of voting rights, $a_i$, is equivalent to delegating votes from the same number of  voting rights proportionally to all voters, including voter i, based on the voting rights holdings that they will exercise in the actual vote.} \\
 
This observation follows from the fact that both abstention and the equivalent delegation leave all voters with the same proportion of the total number of voting rights that are actually voted.

The optimal weights under the Optimal Weighting Theorem, which requires the assumption of independent competencies, are attainable by partial
abstention:\\
 
 \textbf{Proposition 3.2.}  \emph{Assume that voter competencies are independent and that there is at least one $j$ such that $p_j > 0.5$. Define:}
 $$R = \sup_i \frac{w_i}{t_i}$$
 \emph{where $w_i$ is the optimal weight for voter $i$ as defined in equation (\ref{optimal_weights}) and $i$ holds $t_i$ total voting rights.  Suppose that each voter abstains from voting on $a_i \le t_i$ of such voting rights, using only $t^*_i = t_i - a_i$ to vote, and that each voter chooses $a_i$ such that:} 
 $$\frac{w_i}{t^*_i} = \frac{w_i}{t_i-a_i} =  R$$
 
 \noindent \emph{Then the probability of a correct collective decision under 
 decision rule $d$ is maximized.} \\
 
 \textbf{Proof.} Note that by Assumption 3, we effectively have $p_i \ge 0.5$ for all i, ruling out negative weights. Define $W = \sum_{i=1}^{N} w_i$. $W > 0$ because $w_i \ge 0 ~\forall ~i$ and $w_j > 0$ for at least one $j$. Define $K = \frac{1}{W}$. Define $w_i^*$ as the effective weights in the actual vote. Then:
 $$w_i^* \equiv \frac{t_i^*}{\sum_{i=1}^{N} t_i^*} = \frac{\frac{w_i}{R}} {\sum_{i=1}^{N} \frac{w_i}{R}} = K w_i$$ 
 where $K = \frac{1}{W}$ is a constant. The effective weights $w_i^*$ are proportional to the optimal weights $w_i$, resulting in identical collective voting outcomes. $\square$\\

 \noindent Intuitively, if voters abstain such that all voters' ratios of their optimal weights to the number of voting rights they will exercise are the same, then the ratio of the voting rights exercised by each voter compared to the total voting rights exercised by all voters will be proportional to the optimal weight for that voter, resulting in the same collective decision as employing the optimal weights directly.
 
Given independent competencies, cognizable probabilities, and the shared epistemic objective, voters can arrive at an collective decision in accord with the Optimal Weighting Theorem with no need for coordination other than knowledge of $R$:\\

\textbf{Corollary 3.3.} \emph{Under the assumptions of Proposition 3.2, if each voter shares the epistemic objective (Assumption 2), knows $R$, and knows  $p_i$, their own competency (Assumption 3), then the best possible collective decision is attainable without further coordination.}\\

\textbf{Proof.} Knowing $p_i$ enables voter $i$ to compute $w_i$, the optimal weight for that voter. Acting in accord with the epistemic objective, each voter will exercise only $t_i^* = \frac{w_i}{R}$ out of the total voting rights, $t_i \ge t_i^*$, that the voter could exercise. Then by Proposition 3.2 the probability of a correct collective decision is maximized. $\square$\\

The assumption that each voter knows $R$ is implausible. That would require knowing the full panel of voting rights, ($t_1, t_2, ..., t_N$) and the full panel of competencies, ($w_1, w_2, ..., w_N$). Fortunately, the next Corollary suggests a way to largely or entirely circumvent this problem. The voters can coordinate on any value $\tilde{R} \ge R$, and, without affecting the efficacy of the mechanism, we can allocate initial votes in any way desired:\\

\textbf{Corollary 3.4.} \emph{The operation of Proposition 3.2 and the result in Corollary 3.3 are unaffected by either or both of the following modifications:}\\

\emph{(i) Reallocating the panel of voting rights from ($t_1, t_2, ..., t_N$) to a new panel, ($\tilde{t_1}, \tilde{t_2}, ..., \tilde{t_N}$), subject to $\tilde{t_i} > 0$ for all $i$ such that $t_i > 0$, resetting $R$ to $R = \sup_i \frac{w_i}{\tilde{t_i}}$ accordingly.}\\

\emph{(ii) Voters coordinate on a value $\tilde{R} \ge R$ rather than $R$.}\\

\textbf{Proof.} Proposition 3.2 applies for any panel of voting rights with entirely positive values if the value of $R$ is calculated based on the voting right levels in that panel. The proof of Proposition 3.2 is unaffected by shifting $R$ to $\tilde{R}$ so long as $\tilde{R} \ge R$. If $\tilde{R} < R$ then it will not be the case that $t_i^* = \frac{w_i}{\tilde{R}}$ for the voter with the highest $\frac{w_i}{t_i^*}$ ratio and possibly for other voters.$\square$\\

This Corollary suggests an approach to address the unrealistic assumption that all voters know $R$. Rather than coordinating on $R$, voters can coordinate on a large number, $\tilde{R}$, that is likely to be greater than $R$. Furthermore, by reallocating all the voting rights equally among all voters, the possibility that $R$ itself might be very large due to the most competent voter having relatively few voting rights is eliminated. There is no requirement that $\sum_{i=1}^{N} \tilde{t_i} = \sum_{i=1}^{N} t_i$.

The primary problem that arises from setting $\tilde{R}$ too small is that highly competent voters or signals will not be able to exercise enough voting rights. The next example indicates, however, that it is easy to pick a large enough $\tilde{R}$ so that the danger of derailing effective collective choice is minimal even if $\tilde{R} < R$, violating condition (i) in Corollary 3.4. Prior to the example we state and prove a Lemma and a Corollary of the Lemma that are necessary for easy and clear exposition of the example.\\

Define $N_{-i} = (1, 2, 3, ..., i-1, i+1, i + 2, ...n)$, a subset of $N$, and define $S^*_{-i} = \displaystyle\min_{A \subseteq N_{-i}, B = N_{-i}/A} \left| \sum_{j \in A} t_j - \sum_{j \in B} t_j \right|$, where $t_j$ is the voting weight for voter $j$. Note that the weights defining the voting rule in this case need not be optimal. $S^*_{-i}$ is the smallest gap between the sum of the weights for a winning and a losing coalition attainable when all voters except voter $i$ arrange themselves on opposing sides of a binary choice. Voter $i$ is \emph{minimally decisive} if $\left|t_i\right| > S^*_{-i}$. If voter {i} is not minimally decisive, then there will be no configuration of voters for which voter {i} alone will be decisive.\footnote{More than one voter may fail to be minimally decisive. For example, suppose that there are three voters with weights 3 each and that there are a large number of other voters whose weights, all positive, total 1. Then none of these other voters is minimally decisive, and, indeed the collective mass of these other voters is not itself minimally decisive. Only the three high weighted voters matter with respect to the collective choice. This analysis and the definition and role of $S^*_{-i}$ derive from \citet[Corollary 1]{Nitzan_1982}.}\\

\textbf{Lemma 3.5.} \emph{If voter competencies are independent and voter $i$ is minimally decisive, then an increase (decrease) in the epistemic competence, $p_i$, of that voter, will increase the probability of making a correct collective decision under voting with fixed weights when the voter's weight, $t_i$, is strictly positive (strictly negative).}\\

\textbf{Proof.} Define a set of winning coalitions for a voting rule with fixed voting weights for each voter: $W = \left\{S \subseteq N | \sum_{i \in S} t_i > 0.5 \sum_{i \in N} t_i\right\}$. Assume that there are no subsets for which $\sum_{i \in S} t_i = 0.5 \sum_{i \in N} t_i$, a reasonable assumption in sufficiently rich environments in which such subsets would be a set of measure zero. The probability, $P$, of a correct collective choice under the weighted voting rule is:
\begin{equation}
P =  \sum_{S \in W} \prod_{j \in S} p_j \prod_{k \notin S} (1-p_k).
\end{equation}
The partial derivative, $\diffp{P}{{p_i}}$ will produce a term for every term in the sum for $P$ because each voter $i$ is either in $S$ or its complement. Suppose first that $t_i > 0$. When the voter is in the complement of S, differentiation will result in a term, $T$, that, without loss of generality, looks like:
$$ T = - \prod_{j \in S} p_j \prod_{k \notin S, k \ne i} (1-p_j).$$
However, if $S$ is a winning coalition and $t_i > 0$, then $S_{+i} \equiv S + \{i\}$ also is a winning coalition. The corresponding term in P is:
$$\prod_{j \in S_{+i}} p_j \prod_{k \notin S_{+i}} (1-p_k) = (-T) p_i.$$
whose partial derivative with respect to $p_i$ is $-T$. Thus, for each term in the partial derivative $\diffp{P}{{p_i}}$ that is negative, there is an offsetting positive term. However, if voter {i} is minimally decisive, there will be at least one positive term in the partial derivative that is not matched by a negative term. To see that, note that if voter $i$ is minimally decisive and $t_i > 0$, then there exists at least one winning coalition, $\bar{S}$ with $i \in \bar{S}$, such that $\bar{S}-\{i\}$ is not a winning coalition. The positive term in the partial derivative that results from differentiating the term in P for $\bar{S}$ will have no corresponding negative term because $\bar{S}-\{i\}$ is not a winning coalition.\\ 
\indent The proof for the case $t_i < 0$ is similar in an obvious way. One starts with voter $i$ in a winning coalition, takes the partial derivative, shows that the resulting positive term is matched with a negative term of the same magnitude that occurs when $i$ is added to the losing coalition before differentiation. Because $t_i < 0$, the losing coalition remains a losing coalition. Then one shows that if voter $i$ is minimally decisive, there will be negative terms in the partial derivative $\diffp{P}{{p_i}}$ not matched by positive terms. $\square$.\\

\textbf{Corollary 3.6.} \emph{If voter competencies are independent and voter $i$ is not minimally decisive, then changes in that voter's epistemic competence, $p_i$, will have no impact on the probability of making a correct collective decision under voting with fixed weights.}\\

\textbf{Proof.} If voter $i$ is not minimally decisive, then every winning coalition that contains $i$ is matched in a one-to-one correspondence to a winning coalition that excludes $i$. Each such pair will result in terms in the partial derivative, $\diffp{P}{{p_i}}$, that are equal in magnitude but of opposite signs. This result is obvious. If a voter never affects the outcome, it does not matter how competent they are. $\square$\\

\noindent Now the example.\\

\textbf{Example 3.7.} Suppose N = 100, that is, there are 100 voters who hold voting rights. Reallocate the number of voting rights for each voter $i$ to $\tilde{t_i} = 1$. Now $R = \sup_i w_i$, the largest optimal voting weight across all voters, and because the number of voters is finite, there will be a voter $m$ such that $R = w_m$. Assume voters do not know the value of $R$. Set $\tilde{R} = 1000$, and assume all voters know $\tilde{R}$ and use it to coordinate their abstentions. If $w_m \le 1000$, then $\tilde{R} \ge R$ and Corollary 3.4 implies that the voting rule remains optimal. If $w_m > 1000$, then $\tilde{R} < R$ and voters will partially abstain in a way that assigns an effective weight of 1000 to voter $m$, which is less than the optimal weight for voter {m}. By Lemma 3.5 and Corollary 3.6, the probability of a correct collective choice in this situation will be a least as large as the reference situation in which voter $m$ actually had an optimal weight equal to $\tilde{R}$, and in which no other voter had a greater optimal weight.\footnote{If voter $m$ is minimally decisive, then Lemma 3.5 applies, and the probability of a correct collective choice will be higher than in the reference situation. If voter $m$ is not minimally decisive, then Corollary 3.6 applies, and the probability of a correct collective choice will be the same as in the reference situation. There are voter profiles in which voter $m$ would not be minimally decisive. For example, if there are an even number of voters all with epistemic competence $p_m$, then there is no case except for ties in which voter $m$ can change the outcome. $\left|w_m\right| = S^*_{-i}$ rather than $\left|w_i\right| > S^*_{-i}$ , which is required by the definition of minimally decisive.} In the reference situation, the probability of a correct collective decision is at least $p_m$. If $w_{m} > \sum_{i \in S, i \ne m} w_i$, voter $m$ will dictate the result, and the probability of a correct collective choice will be exactly $p_m$. If, instead, $w_m < \sum_{i \in S, i \ne m} w_i$, the probability of a correct collective choice will be greater than $p_m$. But $p_m$ for $w_m$ = 1000 is very high. $1- p_m \approx 5 \times 10^{-435}$. If we had chosen $\tilde{R} = 20$, a much smaller number, the probability of an incorrect collective choice would be no greater than $2.1 \times 10^{-9}$, about two out of a billion.\\
\indent It is clear that for any $\epsilon > 0$, no matter how small, we can pick an $\tilde{R}$ sufficiently large that $1 - p_m \le \epsilon$. In the case of $R > \tilde{R}$, the probability of making a correct collective choice will be in the interval $[1-\epsilon, 1]$ and we will fall short of maximizing that probability by at most $\epsilon$. Because we do not fall short at all when $R \le \tilde{R}$ we are within $\epsilon$ of the maximum in all cases. We conclude that we can approximate attainment of the maximum probability of a correct collective decision to any precision desired. $\square$\\

We now shift from the independent competencies case (Assumption 5A) to the case in which there are dependencies (Assumption 5B). Based on Assumptions 3 and 4, each voter can cognize their information as a set of independent signals that are elements of a canonical list that contains all such signals across all voters. The coordination problem is evident from Example 2.1, which had 3000 voters with optimal weight $w_i =0.1$ and one voter with optimal weight $w_i = 1$. If the 3000 voters do not have independent competencies but, in fact, are perfectly correlated, then they should carry a single weight of $0.1$ rather than, collectively, $300$, which would be appropriate in the independence case. In order to correct for this difficulty, each of the 3000 voters needs to know that the signal implicit in their competency is shared with 2999 other voters. If that is the case, these voters can simply abstain on all but 1/3000 of each of their votes, with the result that collectively their signal would receive the proper weight of $0.1$ versus the independent signal received by the more competent voter. 

This discussion suggests the following assumption as a vehicle for addressing the dependency case:\\

\textbf{Assumption 7 (Cognizable Signal Numeracy):} \emph{For each independent signal that a voter receives, the voter knows how many other voters received the same signal.}\\

\noindent This assumption is both strong and qualitatively different than the assumptions we made in the independence case. In that case, each voter only had to have knowledge of their own epistemic competence and the ability to coordinate with other voters on a single large number as a basis for determining the optimal degree of abstention. In the dependency case, voters must know about other voters' signals in order to achieve an optimal collective choice in a decentralized manner.

Based on Assumption 3, a single governing party who received all of the signals on the canonical list could use the Optimal Weighting Theorem to reach the best possible collective result. Under Assumption 3, that party could compute the probability of reaching the correct collective result based on each signal by itself. Then the party could assign optimal weights to each signal based on these probabilities, create a vote based on the signals received, and maximize the probability of a correct collective decision. 

The same outcome as this centralized approach can be achieved in a decentralized manner. Each voter does the same computation of probabilities for the set of signals which that voter receives. The voter computes the appropriate weights but then divides each weight by the number of voters who receive that signal, a number known to each voter by Assumption 7. There is an additional step. As in the independence case, each voter must abstain on a proportion of that voter's voting rights in order to take into account the fact that the voter only receives a subset of the signals on the canonical list. We complete the description and state the result as a Proposition:\\

\textbf{Proposition 3.8.}  \emph{Assume cognizable probabilities, information decomposition, and cognizable signal numeracy (Assumptions 3, 4, and 7). Assume that voter competencies are dependent because voters each receive a set of independent signals, each of which may be received by some distinct subset of other voters. Define $C$ as the canonical set of all the independent signals received by at least one voter. Define $C_j \subseteq C$ to be the set of independent signals received by voter $j$. Without loss of generality, assume that each voter has one infinitely-divisible voting right on which the voter may partially abstain to any degree. Suppose each voter $j$ uses the following algorithm to decide how to vote and what proportion of the unit voting right to exercise:}\\

\emph{(i) Voter $j$ observes $C_j$, the set of all independent signals received by that voter, and computes based on each signal, $i \in C_j$, the probability, $p_i$, of making a correct collective choice based only on that signal.}\\

\emph{(ii) Voter $j$ computes an optimal weight, $w_i$, for each signal $i \in C_j$ and then divides that weight by the total number of voters who receive that signal to arrive at a final set of adjusted weights, $\{w_i^a ~|~ i \in C_j\}$}.\\

\emph{(iii) Voter $j$ allocates its voting rights in proportion to the weights, with each signal $i \in C_j$ receiving an allocation of $r_{ij}$ voting rights.}\\

\emph{(iv) For each signal $i \in C_j$, voter $j$ abstains on $a_{ij}$ of the $r_{ij}$ voting rights, voting only $r_{ij}^* = r_{ij} - a_{ij}$ voting rights, choosing $a_{ij}$ such that:}
$$\frac{w_i^a}{r^*_{ij}} = \frac{w_i^a}{r_{ij}-a_{ij}} =  R = \sup_{i \in C_j,~ j \in N} ~\frac{w_i^a}{r_{ij}}$$

\emph{(v) Voter $j$ exercises the voting rights accordingly, netting out offsetting votes if different signals suggest a different choice between the two alternatives.}\\

\noindent \emph{If all voters follow this algorithm, then the probability of a correct collective decision under decision rule $d$ is maximized.}\\ 
 
 \textbf{Proof.} The proof is identical to the proof of Proposition 3.2 in the independent case, substituting independent signals for independent voters, and noting that signals received by more than one voter aggregate to the equivalent of one signal with the correct weight. To see this aggregation point, consider a signal $i$ that is received by $K$ voters. The number of voting rights that each voter $k$ commits to the signal is:
$$r^*_{ik} = \frac{w_i} {K R}$$
\noindent noting that by Assumption 3, all voters compute the same value for the optimal weight, $w_i$, for any given signal $i$. The aggregate for signal $i$ will be:
$$\sum_{k=1}^{K} r^*_{ik} = \frac{w_i}{R}.$$
As a consequence, the weights for the signals using the algorithm will be proportional to the optimal weights with the same proportionality constant, and by the Optimal Weighting Theorem, the probability of making a correct collective choice is maximized. $\square$. \\

The result is the same as if a single governing party received all the independent signals and made the collective decision using optimal weights for each signal. The coordination steps $(ii)$ - $(iv)$ allow the voters to achieve the same result in a decentralized setting. If $R$ is not common knowledge, the same approximation approach applicable in the case of independent competencies is available:\\

\textbf{Corollary 3.9.} \emph{If all the assumptions of Proposition 3.8 apply except that $R$ is not common knowledge among the voters, voters can achieve an approximately optimal result of any degree of accuracy with respect to maximizing the probability of a correct collective decision by coordinating on sufficiently high substitute value, $\tilde{R}$, rather than $R$ in the manner described by Corollary 3.4 and Example 3.7.}\\
 
\textbf{Proof.} The Corollary follows immediately from Proposition 3.8, Corollary 3.4, and the reasoning in Example 3.7. $\square$. \\

\section{Direct Participation}\label{participation}

As discussed in the Introduction, low participation rates have been a major concern for DAOs. ``Democratic" ideals and the desire to avoid de facto centralized control both create an impetus to increase direct participation. From an epistemic standpoint, however, the desirability of increasing direct participation is much more nuanced, and it depends heavily on the details of the composition of voting rights across voters. In this subsection, we explore the nuances.

There is an extensive literature relevant to the epistemic consequences of direct participation, particularly in the context of one-person-one-vote systems. We do not attempt to discuss that literature or its implications comprehensively, but instead focus on some basic results and examples that illustrate major contours of the epistemic implications of direct participation as well as aspects that are particularly pertinent to the discussion of transfer delegation in the next section.

We follow \citet{Ben_Yashar_2000} by defining the voting group's \emph{competence structure} as $\underline{p} = (p_1, p_2, ..., p_N)$ where we order the voters such that $p_i \ge p_j$ for $i > j$. We also define the corresponding \emph{optimal weight structure} as $\underline{w} = (w_1, w_2, ..., w_N)$ and the \emph{voting rights structure} as $\underline{v} = (v_1, v_2, ..., v_N)$, which expresses the actual weights under the voting rule normalized such that 
$$\sum_{i=1}^{N} v_i = \sum_{i=1}^{N} w_i = W.$$
\noindent I.e., if each voter $i$ has $t_i > 0$ voting rights, then we set:
$$v_i = \frac{t_i}{\sum_{j=1}^{N}t_j} W.$$
Given this normalization, the voting rule is the one specified by the Optimal Weighting Theorem if and only if $\underline{v} = \underline{w}$. One-person-one-vote corresponds to $v_i = 1/W$ $\forall~ i$, which we denote as the voting rights structure $\underline{v}^{1P1V}$. Note that as in the previous sections, voters may be delegates with voting rights acquired by transfer delegation. Consequently, the results in this section speak directly to the methods of transfer delegation discussed in the subsequent section.

We begin on a positive note, delineating an environment in which increased direct participation is never harmful and often quite valuable with respect to making the correct collective decision:\\

\textbf{Proposition 4.1.} \emph{In the independence case, if both present and potential voters exercise their voting rights using the weights specified by the Optimal Weighting Theorem, then adding a new voter, $j$, with $p_j\ge 0.5$ weakly increases the probability of making the correct collective choice, with the increase being strictly positive if and only if the new voter is minimally decisive after being added.}\\

Voters with $p_j = 0.5$ are never minimally decisive. Adding them neither increases nor decreases the probability of making a correct collective decision. The same is true for voters with $p_j > 0.5$ who are not minimally decisive after being added. 

Before executing the proof of Proposition 4.1, we state a Lemma that is useful in that proof as well as later proofs in the paper.\\

\textbf{Lemma 4.2 (\citet{Nitzan_1982}).} \emph{In the independence case in which voter $i$ has optimal weight $w_i$, consider two complementary coalitions of voters: $S \subseteq N$ and $S/N$. Then:
$$ \prod_{j \in S} p_j \prod_{k \in N/S} (1-p_k) \lesseqgtr  \prod_{k \in N/S} p_k \prod_{j \in S} (1-p_j)$$
is equivalent to
$${\sum_{j \in S} w_j} \lesseqgtr {\sum_{k \in N/S} w_k}.$$}\\

\textbf{Proof.} This result is part of the proof of the Optimal Weighting Theorem in \citet[Theorem 1, p. 293]{Nitzan_1982}. 
$$ \prod_{j \in S} p_j \prod_{k \in N/S} (1-p_k) \lesseqgtr  \prod_{k \in N/S} p_k \prod_{j \in S} (1-p_j)$$
is equivalent to:
$$\frac{\displaystyle\prod_{j \in S} p_j}{\displaystyle\prod_{j \in S} (1-p_j)} \lesseqgtr \frac{\displaystyle\prod_{k \in N/S} p_k}{\displaystyle\prod_{k \in N/S} (1-p_k)}$$
which reduces to:
$${\prod_{j \in S} \frac{p_j}{(1-p_j)}} \lesseqgtr {\prod_{k \in N/S} \frac{p_k}{(1-p_k)}}$$
which is equivalent to:
$${\sum_{j \in S} w_j} \lesseqgtr {\sum_{k \in N/S} w_k}.$$
$\square$\\

This Lemma is the core element of the proof of the Optimal Weighting Theorem in \citet{Nitzan_1982}. Because it adds insight for what follows, we explain why in the rest of this paragraph. A voting rule is completely determined by specifying which subset in each pair of complementary subsets $(S \subseteq N, N/S)$ is a winning coalition. All subsets are a member of one and only one such pair. For each subset $S$, define a \emph{pattern probability}, $q(S)$:
$$q(S) = \prod_{j \in S} p_j \prod_{k \in N/S} (1-p_k).$$
$q(S)$ is the probability that all members of $S$ vote for the correct alternative while all members of $N/S$ vote for the incorrect alternative. The left-hand side of the first inequality in the Lemma is $q(S)$, and the right-hand side is $q(N/S)$. Because any pattern can occur, $\sum_{S \subseteq W} q(S) = 1.$ Furthermore, the probability of making a correct collective decision under a particular voting rule is the sum of the pattern probabilities associated with each winning coalition under that rule as indicated by equation (2) in the proof of Lemma 3.5. It is obvious that the best possible voting rule will be the one which chooses the winning coalition from each pair of complementary coalitions that has the highest pattern probability. Lemma 4.2 indicates that a voting rule based on choosing the subset in each pair $(S \subseteq N, N/S)$ with the highest sum of optimal voting weights under the Optimal Weighting Theorem accomplishes exactly that result.\\

\textbf{Proof of Proposition 4.1.} Suppose there are $K$ voters in the DAO, all exercising their voting rights for each vote using optimal weights, $w_i$, for each voter $i$ based on the Optimal Weighting Theorem. Add a new voter $j$ for whom $p_j \ge 0.5$ and thus $w_j \ge 0$ for the vote under consideration. Consider two approaches. First, the new voter is a assigned a zero weight, which is equivalent to precluding that voter from participating because with zero weight they can have no impact on the outcome. Alternatively, the new voter is assigned the optimal weight, $w_j$, for that vote. It follows from the Optimal Weighting Theorem that the second approach does not decrease the probability of making the correct collective choice and may increase it if $w_j > 0$. I.e., adding voter $j$ weakly increases the probability of making a correct collective decision.\\
\indent Claim: the probability of making a correct collective decision increases if and only if the added voter, $j$, is minimally decisive after being added. Consider the structure before adding $j$. This structure consists of a set $W$ of winning coalitions, and for each winning coalition $S \subseteq N$, the complement of $S$ in $N$, $N/S$, is a losing coalition. All coalitions belong to one and only one such pair. Because the decision rule weights voters using optimal weights, if S is a winning coalition, then $\sum_{i\in S} w_i > \sum_{i \in N/S} w_i$. The probability, $P$ of a correct collective choice is:
$$P =  \sum_{S \in W} \prod_{i \in S} p_i \prod_{i \notin S} (1-p_i)$$
where each term is the probability of every member of a particular winning coalition voting for the correct choice and every member of the complementary losing coalition voting for the inferior alternative.\\
\indent For each pair consisting of a winning coalition, $S_W$, and a losing coalition, $S_L = N/S_W$, prior to adding voter $j$, define the following:
$$B= \prod_{i \in S_W} p_i \prod_{i \in S_L} (1-p_i).$$
$$C= \prod_{i \in S_L} p_i \prod_{i \in S_W} (1-p_i).$$
$$X= p_j B = p_j \prod_{i \in S_W} p_i \prod_{i \in S_L} (1-p_i).$$
$$Y= (1-p_j) B = (1-p_j) \prod_{i \in S_W} p_i \prod_{i \in S_L} (1-p_i).$$
$$Z= p_j C = p_j \prod_{i \in S_L} p_i \prod_{i \in S_W} (1-p_i).$$
$B$ is the term in $P$ corresponding to $S_W$ being a winning coalition before adding voter $j$, the baseline case. If voter $j$ is not minimally decisive, then both $S_W + \{j\}$ and $S_W$ are winning coalitions after adding $j$. As a result, after adding $j$, the terms $X$ and $Y$ now both appear in the sum for $P$ instead of $B$. But these terms add up to $B$. As a result, $P$ does not change. Suppose instead that $j$ is minimally decisive. Then there is at least one coalition pair consisting of some $S_W$, a winning coalition, and $S_L = N/S_W$, the complementary losing coalition, for which adding voter $j$ to $S_L$ transforms it into a winning coalition. In this case, after adding $j$, both $S_W + \{j\}$ and $S_L + \{j\}$ are winning coalitions, and the terms $X$ and $Z$ replace $B$ in the sum comprising $P$.\\
\indent It remains to prove that $X + Z = p_j (B + C) > B$. This statement is equivalent to $p_j C > (1-p_j) B$, which reduces to: $Z > Y$. By Lemma 4.2, $Z > Y$ if and only if:
$$\sum_{i \in \{S_L +\{j\}\}} w_i > \sum_{i \in S_W} w_i.$$

This inequality is true because $S_L + \{j\}$ is a winning coalition and therefore must have a higher sum of optimal weights compared to its complement, $S_W$. Considering all ``original" pairs of winning and complementary losing coalitions before adding voter $j$, $P$ will increase with respect to every original pair for which voter $j$ can shift the losing coalition to a winning coalition by joining it, and $P$ will remain unchanged with respect to the other pairs. If voter $j$ is minimally decisive, there must be at least one original pair for which $j$ can join a losing coalition and transform it into a winning coalition. $\square$.\\

\noindent The same result applies in the dependence case under Assumption 4 if voters can coordinate to separate out the underlying independent signals and collectively give them the proper weight, for example, using the method delineated in Proposition 3.8:\\

\textbf{Corollary 4.3.} \emph{In the dependence case under Assumption 4, if both present and potential voters can coordinate to give each independent signal $j$ the optimal collective weight $w_j$ based on the fact that use of that signal by itself results in a $p_j$ probability of making the correct collective choice, then a new voter will increase the probability of a correct collective choice if and only if the new voter observes one or more  independent signals not observed by any of the existing voters and these signals are collectively minimally decisive after being added to the previous set of independent signals. If the new voter does not observe such a set of signals, then adding the new voter will have no impact on the collective choice.}\\

\noindent The proof is identical to the proof of Proposition 4.1 substituting independent signals for voters with independent epistemic competencies. In the rest of this section, we will state many of the results in the context of the independent competencies case with the understanding that the same results hold if Assumption 4 applies and there is a coordination technology to extract and properly weight independent signals so that they play the same role as independent voters. At some points, however, we will explicitly address the dependencies case.

Define the following:\\

\textbf{Optimal Epistemic Environment}: \emph{An optimal epistemic environment exists if the voting rule and voter coordination result in the optimal voting weights according to the Optimal Weighting Theorem for either voters with independent competencies or in the dependence case under Assumption 4 for the canonical list of independent signals.}\\

Proposition 4.1 and Corollary 4.3 indicate that in the optimal epistemic environment, direct participation by new voters who bring new information to the governance mechanism weakly increases the probability of making correct collective decisions and causes no harm when new voters arrive who do not. Outside this environment, the epistemic impact of increased direct participation becomes very uncertain and can be extremely negative. Although the possibility of negative epistemic results is more general, we identify three specific dangers that have high practical relevance and also play a major role in the subsequent section that discusses transfer delegation. 

A large number of scholars have examined one-person-one-vote majority voting from an epistemic viewpoint. When voters have independent competencies, this voting system only satisfies the Optimal Weighting Theorem if all voters have the same epistemic competence, $p_i = p ~\forall~ i$. In that case, consonant with the increasing reliability property of Condorcet's Jury Theorem, adding new independent voters with the same level of competence increases the probability of making the correct collective choice. 

When independent voters have heterogeneous competencies, the situation is quite different. It is no longer the case that $\underline{v}^{1P1V} = \underline{w}$. Guaranteeing that adding independent voters with $p_i > 1/2$ appears to require that the added voters have very high epistemic competence compared to the existing voters. {\citet{Ben_Yashar_2017}} examine majority voting (one-person-one vote) when there are an odd number of voters. They show that a sufficient condition \label{pair} for adding two new independent voters, $i$ and $j$, to increase the probability of a correct collective decision is  $w_i + w_j > w_1$. The sum of the optimal weights for the two added voters must exceed the optimal weight of the most competent voter in the previous set. 

Extending the analysis in {\citet{Ben_Yashar_2017}}, we obtain the following Corollary:\\

\textbf{Corollary 4.4.} \emph{Assume majority rule with a set of voters $N$ where $n = |N|$ is odd and  $p_i \ge 0.5$ $\forall~ i$. Let $S$ denote a subset of $N$ and $N_{x}$ the set of all subsets of $N$ of size $x$. Define:}
$$
\mathrm{A}=\sum_{S \subseteq N_{\frac{n-1}{2}}} \prod_{k \in S} p_{k} \prod_{k \in N \backslash S}\left(1-p_{k}\right)
$$
$$\mathrm{B}=\sum_{S \subseteq N_{\frac{n+1}{2}}} \prod_{k \in S} p_{k} \prod_{k \in N \backslash S}\left(1-p_{k}\right)$$
$$\phi=\frac{B}{A}.$$
\emph{Then $\phi \ge 1$ with equality only in the case $p_k = 0.5$ $\forall ~ k$. Consider adding two new voters with competencies $p_i$ and $p_j$. Then:}\\

\noindent \emph{(i) The probability of making a correct collective choice increases (decreases) if and only if $w_i + w_j > (<)~ ln(\phi)$.}\\

\noindent \emph{(ii) If $p_i = p_j = 0.5$, then adding these voters will decrease the probability of making a correct collective choice unless that probability is $0.5$ because $p_k = 0.5$ $ \forall ~ k \in N$.}\\

\noindent \emph{(iii) If at least one voter $k \in N$ has $p_k > 0.5$ and the two new voters, $i$ and $j$, each have some epistemic competence, that is $p_i > 0.5$ and $p_j > 0.5$, then there will be a range of values defined by $w_i + w_j < \phi$ for which adding these two voters will decrease the probability of a correct collective choice.}\\

\noindent \emph{(iv) For any $\epsilon > 0$, it is possible to reduce the probability of a correct collective decision below $0.5 + \epsilon$ by repeatedly adding new voters with $p_k > 0.5$.}\\

\textbf{Proof.} We have used an identical setup as \citet{Ben_Yashar_2017}, and $A$ and $B$ are exactly the same as stated in their paper. They show that the change in the probability of a correct collective decision from adding two new voters, $i$ and $j$, to $N$ is $\Delta = p_{i} p_{j} \mathrm{A}-\left(1-p_{i}\right)\left(1-p_{j}\right) \mathrm{B}$ where A is the probability that exactly $\frac{n-1}{2}$ voters in the original set make the correct choice and B is the probability that exactly $\frac{n+1}{2}$ of such voters make the correct choice. $A > 0$ and $B > 0$.\footnote{The ratio $\phi = B/A$ between them depends on the precise competence structure of the original set of voters. Thus, the necessary and sufficient condition stated as (i) in the Corollary is not general but depends on that structure. In contrast, the sufficient condition in \citet{Ben_Yashar_2017} that $w_i + w_j > w_1$ implies a higher probability of a correct collective decision is general, independent of the precise competence structure in any given case.} Given $\phi = B/A$, straightforward arithmetic establishes that the following three equations are equivalent:
\begin{align*}
  \Delta & > (<) ~ 0 \\
  p_ip_j  & > (<) ~ (1-p_i)(1-p_j)~ \phi\\
  w_i + w_j  & > (<) ~ ln(\phi)
\end{align*}
It is easy to see that $B \ge A$. $\binom{n}{\frac{n+1}{2}} = \binom{n}{\frac{n-1}{2}}$ for $n$ odd, and there is the following bijective correspondence between the terms of $B$ and $A$: Choose a subset $S  \subseteq N_{\frac{N+1}{2}}$ and some $x \in S$. Consider the term in B arising from $S$ and $N/S$:\\
$$ \prod_{k \in S} p_{k} \prod_{k \in N/S}\left(1-p_{k}\right)=p_x ~ \prod_{k \in S/\{x\}} p_{k} \prod_{k \in N/S}\left(1-p_{k}\right).$$
\noindent For each such term in B, there is a single corresponding term in A equal to:\\
$$ \prod_{k \in S/{x}} p_{k} \prod_{k \in \{N/S +\{x\}\}}\left(1-p_{k}\right)= (1-p_x) ~ \prod_{k \in S/\{x\}} p_{k} \prod_{k \in N/S}\left(1-p_{k}\right).$$
\noindent Because $p_k \ge 0.5$ $\forall~ k$, the term from B is greater than or equal to the term from A with equality only when $p_x= 0.5$. The same is true for the exhaustive set of all such pairs of terms from $B$ and $A$. Thus, $B \ge A$ with equality only when $p_k = 0.5 ~ \forall ~ k \in N$. (ii) and (iii) follow immediately from the sign of $\phi$ under the respective assumptions about competence structure for those two parts of the Corollary.\\
With respect to (iv), assume the opposite. That is, there is a level $P_L > 0.5$ that is a lower bound for the probability of a correct collective decision that is impervious to adding more voters with $p_i > 0.5$. Because this lower bound is greater than $0.5$, at the lower bound it must be the case that there is at least one current voter $k$ with $p_k > 0.5$. Otherwise, $P$, the probability of making a correct collective decision, would be $0.5$. But the presence of at least one voter $k$ with $p_k > 0.5$ implies that $\phi >1$ and $ln(\phi) > 0$. By (iii), it is possible to pick two new voters $i$ and $j$ with $p_i > 0.5$ and $p_j > 0.5$ such that $P < P_L$, a contradiction. $\square$.\\

Result (iv) in the Corollary exemplifies the first danger:\\

\noindent \textbf{The Flooding Danger}: Outside of the optimal epistemic environment, there is the danger that adding voters with low levels of competence will significantly lower the probability of making a correct collective decision even if: (1) each of the added voters have independent competencies with respect to each other and all existing voters; and (2) each added voter $k$ has $p_k > 0.5$.\\

In the optimal epistemic environment, the flooding danger is not present because added voters receive optimal weights, which means appropriately low ones for added voters with low competencies. Instead of potential detriment, adding independent low competence voters each with $p_i > 0.5$ weakly increases the probability of making a correct collective choice. These added voters are introducing fresh, valuable information which potentially enhances the ability of the DAO to steer in the right direction.

The Flooding Danger comes strongly into play under policies that attempt to enhance direct participation by drawing in voters who are either ``rationally ignorant" or ``rationally ambivalent." These voters have strong reasons not become informed about the issues relevant to voting, and, consequently, are likely to have low competence. Many of them may be ``coin flip" voters with $p_i = 0.5$ and optimal weights $w_i = 0$. In that case the Optimal Weighting Theorem clearly indicates that the best strategy is not to vote, a strategy that is equivalent to delegating their voting rights to the voters who do participate (Observation 3.1 above). Efforts to induce such voters to participate may have quite negative epistemic consequences.\footnote{\citet{Paroush_1997}, who reaches conclusions related to part (iv) of Corollary 4.4, includes the phrase ``stay away from fair coins" in the title of his article and notes that ``one cannot even exclude the counter-intuitive case where all $p_i > 1/2$ and yet lim $\pi = 1/2$ when the team size tends to infinity." Stating this point more formally in a way that directly follows from his analysis: \emph{Given any $\epsilon > 0$, majority rule may result in a probability of a correct collective choice less than $0.5 + \epsilon$ despite having a countably infinite supply of independent voters all of whom have $p_i > 0.5$.} He notes that for a collection of $n$ voters with $w_1 > \sum_{i=0}^{n} w_i$, the Optimal Weighing Theorem implies that voter 1 making the decision unilaterally (``expert rule") is superior to majority rule. \citet[Corollary 2(a)]{Nitzan_1982}. It follows that majority rule among the remaining $n-1$ voters will result in a probability of a correct collective choice less than $p_1$. Given an $\epsilon > 0$, set $p = 0.5 + \epsilon$ and a corresponding optimal weight, $w = \frac{p}{1-p}$. Construct a countably infinite set of voters with an optimal weights given by $w_i = \frac{1}{2^i} w$. Because $\sum_{i=1}^{\infty} w_i = w$, majority rule will result in the probability of a correct collective choice less than $p$.}

There are many reasons to expect that a large proportion of token holders and even possibly a significant  proportion of delegates will fall into the rationally ignorant category with respect to many votes. Except for voters holding a significant portion of the economic (non-voting) rights inherent in the total token supply, the benefits of any effort to become an informed voter will accrue mostly to other token holders who are in effect free riders.\footnote{\citet[pp. 239-240]{Khanna_2022} describes this collective action problem in the corporate setting with reference to some of the literature. \citet[p. 26]{Reyes_2017} appear to be the first to note the presence and importance of the same problem in the DAO setting.} Delegates may have very limited economic rights even though they have substantial voting rights, making the question of how to motivate them important. Regardless of the proportion of economic ownership, a voter who holds a relatively modest share of the total voting rights will be very unlikely to be decisive in any given vote, and voting is costly in terms of time and other resources. Portfolio investors may hold a substantial proportion of some voting rights. Many of these investors will hold a large number of different tokens with each forming a relatively small portion of their portfolios, a factor that favors rational ignorance with respect to any one token. In addition, portfolio investors, especially if they engage in indexing strategies, may tend to be rationally ambivalent. They may hold competing tokens so that efforts on behalf of one token will create gains on that token that are offset by losses on competing tokens also held in the portfolio. Broad index holdings tend to cancel out idiosyncratic risks such as the competitive offsets just described, leaving only coverage of economy-wide factors.

So far we have been looking at the case in which all voters are independent. As mentioned the same analysis applies if we have properly unpacked dependencies into independent signals, which then play the part of independent voters in the analysis. A second danger arises when we fail to do so, leaving dependencies in place with no adjustments and putting us outside of the optimal epistemic environment which requires the unpacking.\\

\noindent \textbf{The Dependency Danger}: If we are not in the optimal epistemic environment because dependencies between voters are not taken into account, there are plausible scenarios in which the probability of making a correct collective decision will be substantially lower even if the voting method assigns optimal weights to each voter $i$ based on $p_i$ and all voters have $p_i > 0.5$.\\

Example 2.1 is a clear and very relevant example of this danger. The example includes 3000 voters with $p_i \approxeq 0.525$, each of whose votes are weighted optimally at $w_i = 0.1$ along with one ``expert" voter, $k$, with $w_k = 1$ based on $p_k \approxeq 0.731$. The expert voter is independent of all the other voters. If the 3000 voters are mutually independent, Condorcet's Jury Theorem comes into play, and the probability of a correct collective decision is approximately $0.997$. But if the 3000 voters are perfectly correlated, then they all vote the same way, and they are decisive based on the weighted voting rule with a joint weight of 300 versus the expert with a weight of 1. The probability of a correct collective decision collapses to approximately $0.525$, worse than expert rule, which would increase this probability to approximately $0.731$. If the independent signals were unpacked appropriately, there would be two: the expert's vote signal with weight 1 and the single signal received by all 3000 voters with weight 0.1. The optimal epistemic environment that properly accounts for dependencies would be restored, and the result would be the same as expert rule, appropriately in this situation. Instead the expert gets washed out and the probability of making a correct collective decision plunges by more than 20 percentage points.

This type of situation is plausible, especially if a system pressures large sets of rationally uninformed voters into participating. Not wanting to devote much time, which may be totally rational, a large group might, for instance, go online and look at the first result that comes up on Google or another widely-used resource. Then they all receive the same signal, resulting in perfect correlation among a large group of voters with potentially very damaging epistemic consequences.

A third danger involves the opposite problem from too much influence from a diffuse set of voters:\\

\noindent \textbf{Epistemic Danger from Whales}\label{whale_problem}: Certain parties may hold a very large portion of the total voting rights either stemming from their own token ownership or from delegation. In approaches such as one-token-one-vote that do not employ optimal weights, these ``token whales" or ``delegate whales" may have too much weight compared to the optimal level and may largely determine decisions, resulting in a significant decrease in the probability of making a correct collective decision. Other independent information will not receive its due, and Condorcet's Jury Theorem phenomena will be less able to exert a positive influence. \\

In many instances, token whales or delegate whales may have high individual competence. In the case of token whales, having a large economic stake creates a strong incentive to be well informed because the token whale can capture a large percentage of the benefit from accumulating expertise. Aside from being well informed, token whales in particular have a strong incentive to use delegation, including abstention as a form of delegation, to move effective voting weights toward the optimal set. And when the required adjustment rests with a single sophisticated party, it is more likely to occur and more likely to occur in the correct direction than when a large number of small holders must act in a coordinated fashion. 

Both the epistemic danger of whales and the possible solutions are illustrated by the following example of one large voting rights holder surrounded by a large group of identical small voting rights holders:\footnote{The example and its solution draws from \citet{Bar_Isaac_2020} who study the situation of a group of independent voters consisting of a single large, highly-competent voting rights holder surrounded by a large number of smaller holders with equal but much smaller competence levels. They formulate optimal abstention policies that address imbalances in both directions: not enough weight for the large holder and not enough weight for the small holders. When there is not enough weight for the large holder, a coordinated abstention response is required from the large mass of small holders, which is much less plausible than an optimal response by the single large holder in the opposite situation in which that holder has too much weight.} \\

\textbf{Example 4.5.} Assume the same framework as Example 2.1 and assume independent competencies across all voters. There are $30,000$ voting rights outstanding. A diffuse mass of $3000$ voters each with optimal weight $w_i=0.1$ ($p_i \approxeq 0.525$) hold one voting right each, and the remaining ``whale" voter with optimal weight $w_i=1$ ($p \approxeq 0.731$) holds $27,000$ voting rights, $90\%$ of the total. Majority voting without abstention will result in the whale voter dictating all decisions. Although the whale is much more competent than each of the individuals in the diffuse mass, as Example 2.1 indicates, aggregation of the independent views of the diffuse mass with the whale when voters are given their optimal weights will raise the probability of a correct collective decision from $73.1\%$ to $99.7\%$. The whale can easily ensure this result by abstaining on all but $10$ of the whale's total holdings of $27,000$ voting rights. Then all parties will have optimal voting weights. The weights total $301 = 1 + 0.1 \times 3000$, and each voter has a number of voting rights equal to $10$ times that voter's weight. Only about $10.03\%$ of the voting rights will be exercised.$\square$\\ 

The example illustrates some of the ambiguities in the concept of participation. In the example, optimal abstention results in the exercise of very few of the total voting rights. Participation is low in that sense. But unless one is willing to assert that intelligent partial abstention makes the large voting rights holder a non-participant, the participation rate in terms of intelligent engagement is $100\%$. More generally, $100\%$ participation in that sense is required to maximize the probability of making a correct collective decision if each voter has valuable information that is independent of all other voters' information. It also may be the case in some situations, which may be quite common, that what appears to be a low participation rate in terms of engagement is optimal. That happens  if a large proportion of potential voters are epistemically ignorant with $p=0.5$ and corresponding optimal weights of $0$. As we have discussed, it may be rational for these potential voters to be in this state of ignorance. Low participation can be optimal whether it is terms of percentage of voting rights exercised or in terms of the proportion of voters participating at least in part.\footnote{One idea is to include an ``abstain" option so that voters can signal engagement without voting for an alternative. Doing so might add to the perceived legitimacy of the choice process. But some of the reasons that motivate rational ignorance such as owning a very small stake in the DAO suggest that many voters may forgo this costly step.}

\section{Transfer Delegation}\label{delegation}

As suggested in subsection \ref{terminology}, considering transfer delegation requires adding an additional dimension to the concept of competence: \emph{Delegative competence} is the ability to delegate voting rights in a way that increases the probability of a correct collective decision. Delegative competence is distinct from epistemic competence, which has been the sole focus in the previous sections. Recall that epistemic competence of voter $i$ is the probability, $p_i$, that the voter \emph{acting alone} will reach the correct collective decision. It is plausible that epistemic and delegative competence are positively correlated to some degree, but they are distinct.  

Delegative competence itself has multiple components. One component is \emph{weight competence}, the ability to ascertain what the optimal weights are for other voters in the independence case or to ascertain independent signals and their associated optimal weights in the case of dependencies. The other components do not lend themselves easily to a taxonomy and definitions. 

An illustration of some of the considerations flows from the following definition: \emph{Weight omniscience} consists of knowing the optimal weights for all of the voters who will participate in the case of independent epistemic competencies or for all the signals that are present if there are dependencies. If there is an individual or entity with weight omniscience, the best possible collective decision is attainable through a two-step transfer delegation process:

\begin{enumerate} [label=\arabic*)]
\item Voters delegate all of their voting rights to the individual or entity with weight omniscience.
\item That individual or entity delegates voting rights to voters (or signals in the case of dependencies) based on their optimal weights, and all of those parties (or signals) vote without delegating further.\footnote{If participation is less than 100\%, the individual with weight omniscience can achieve the best possible collective decision conditional on the restricted voting pool.}
\end{enumerate}

\noindent It is possible to achieve the same result through abstention. The omniscient individual or entity instructs all the other voters how many voting rights to exercise based on a calculation using optimal weights and the value $R$ as described in Proposition 3.2.

The added element of delegative competence under both of these approaches is that the voters must know that a particular individual or entity has weight omniscience. To know that, the voters must be able to assess the delegative competence of other voters, which goes beyond weight competence, the ability to discern the optimal weights of other voters. The situation becomes even more complex in a multi-step transfer delegation. The voters then need to know how competent other voters are at identifying the delegative competence of other voters. If many steps are permitted, the situation becomes even more complicated as voters must be able to assess how delegative competencies play out in a lengthy multi-step chain.

The example of what can be accomplished if there is an individual or entity with weight omniscience illustrates another aspect. To exploit the capabilities of this entity or individual purely through transfer delegation as opposed to an approach based on partial abstention, the delegative process must have at least two steps. The omniscient voter receives voting rights in the first step but needs the second step to optimally re-delegate them. 

More generally, when transfer delegation is the choice method, an important structural parameter is the number of delegative steps that are allowed. We  consider three separate cases:\\

\textbf{Assumption 8A (Single-step transfer delegation):} \emph{Voters delegate in a single step. After that step is completed, no further transfer delegation is permitted.}\\

\textbf{Assumption 8B (Unlimited-step transfer delegation - liquid dem-ocracy):} \emph{Voters delegate in a series of rounds. Following each round, voters observe the outcome of that round and may change their transfer delegations accordingly. The rounds continue until no voters want to change their transfer delegations in response to the previous round.}\\

\textbf{Assumption 8C (Multiple-step transfer delegation):} \emph{Voters delegate in a limited and fixed number of rounds. After each round, voters may change their transfer delegations after observing the outcome of the previous round. After the final round, no further transfer delegation is permitted.}\\

\noindent We also will consider \emph{sortition}, a method of delegation that operates by choosing a random sample of voters who then vote as a group to make the collective decision. Sortition can operate on a one-voter-one-vote basis or can add members based on metrics such as token holding, an approach which makes representation more likely for parties weighted more heavily by the chosen metric. Under this second approach, parties may end up with multiple votes within the chosen decision group.

All four of these methods of delegation are accompanied by diverse important nuances and a substantial literature. A full discussion of the epistemic properties of each method might require multiple lengthy papers. We limit ourselves here to a few major insights that follow from the framework and results we have derived so far, devoting a separate subsection to each method. Before doing so, we discuss a single delegator's decision and relate it to some practical developments. And after considering each of the methods, a final subsection examines the choice of the \emph{delegate slate}, the set of possible delegates available to voters.

\subsection{List-based Transfer Delegation}\label{list_based}

A delegator uses \emph{list-based transfer delegation} when the delegator chooses to allocate voting rights to more than one delegate from a list set by the delegator, with rules specifying how to move forward if some delegates are unavailable. The list may include the delegator. List-based transfer delegation is a topic of current significance.\footnote{For instance, Graham Novak, a pioneer in introducing this kind of delegation, has been active in its potential development. Novak called a first version, ``Rank-Choice Delegation," which he described as follows: ``[E]ach voter delegates their votes to an ordered list of preferred delegates. The voting delegate who is highest on a voter's list [and available] will add the voting power of the voter to their vote. [There is a] greater probability that someone's interests are represented in each vote since there a many fallbacks." \citet{Novak_2024}.} 

Consider the implications of the Optimal Weighting Theorem for the transfer delegation decision of a single individual or entity assuming one-step transfer delegation and no ability to coordinate with others. In this case, if potential delegates have independent competencies, the Optimal Weighting Theorem dictates that delegators should spread their voting rights across multiple delegates, including possibly retaining some voting rights as a ``self-delegation," using weights based on the probabilities that each delegate will vote in accord with the objectives of the delegator. This approach is superior to simply allocating all voting rights to the single delegate with the highest epistemic competency with respect to the delegator's objectives if any of the other delegates has a positive optimal weight. 

Because optimal weights are absolute quantities in the independence case, they do not need to be recalculated if some delegates are not available. However, in the case of unavailability of some delegates, it is necessary to reallocate the voting rights proportionally based on the remaining weight structure. If there are dependencies, re-weighting based on lack of availability of some delegates is more complex.\footnote{Suppose there are 11 independent signals and 10 delegates. Assume that the first ten signals each are received by only one of the 10 delegates, but the eleventh signal is received by both delegate 1 and delegate 2. Suppose also that the eleventh signal is the one with by far the highest optimal weight and that all the other signals have equal optimal weights. In this scenario, if all delegates are available, delegates 1 and 2 will receive equal voting rights and much higher levels than the other delegates if the delegator acts optimally. Suppose that the situation changes so that all delegates except delegate 1 are available. Re-weighting by just using the weights for the other nine delegates in the full availability situation is not optimal. It would underweight the eleventh signal.}

These results with respect to list-based delegation apply independent of whether voters share the epistemic objective of reaching a correct collective decision or have instrumental objectives such as identifying and promoting their own preferences with respect to the collective choice. The latter situation is particular pertinent if there is no correct collective choice and the collective decision process aggregates preferences instead of information. When preferences are the focus, voters will want to delegate to parties able to discern and implement their preferences as in the main model in \citet{Bloembergen_2019}. The Optimal Weighting Theorem applies for voter $j$'s delegation decision except that each probability $p_i$ is the probability that delegate $i$ correctly discerns voter $j$'s preferences with respect to the binary collective decision rather than the probability that delegate $i$ will make the correct collective choice between the two alternatives in the environment of Assumption 1 in which there is a correct  collective answer.

\subsection{Transfer Delegation in One Step or a Finite Number of Steps}

Proposition 3.2 establishes that, in the case of independent competencies, it is always possible to create a pattern of abstention that results in optimal weights and the highest probability of making a correct collective decision. It is obvious that transfer delegation can achieve the same result. Voters with too many voting rights can delegate them to voters with too few in such a way that the proportion of voting right holdings exactly matches the proportions derived from the optimal weights.\footnote{If some voters have negative optimal weights, that is, $w_i < 0$, then we rely on Assumption 3 that voters know their optimal weights and also on Assumption 2 that voters share the epistemic objective of maximizing the probability of a correct collective decision. Based on these assumptions, negative optimal weight voters will vote the opposite of their inclinations on the binary choice at hand, and the transfer delegation participants can treat them as if they had positive optimal weights equal to $|w_i|$.} However, the coordination problems are much more difficult and the information requirements are much more stringent for transfer delegation to be successful than for implicit delegation through abstention. In fact, optimal transfer delegation in a decentralized environment is very difficult to attain even assuming independent competencies regardless of the number of steps. We begin with the following Proposition, stated in terms of a single transfer delegation step:\\

\textbf{Proposition 5.1} \emph{Consider a single voter who is deciding: (1) whether to delegate at all by explicit transfers of voting rights; and (2) if transfer delegation is indicated, to whom to delegate and in what amounts. Assume:}
\begin{enumerate} [label=(\Alph*)]
\item \emph{Independent competencies, and all epistemic competencies are positive.}
\item \emph{All other voters have delegated, and the net result of their transfer delegations is known by the voter.}
\item \emph{The voter knows which parties are in the set $\bar{V}$ of other voters or delegates who will actually exercise at least some of their voting rights and knows the number of voting rights $\bar{t_j}$ that each such voter $j \in \bar{V}$ will exercise.}
\item \emph{The voter knows the optimal voting weights for all the participants in the set $\bar{V}$.}
\end{enumerate}
\emph{If voter $i$ is minimally decisive given the profile of voting rights
 exercise by the other active participants, $\bar{t} = \left( \bar{t_1}, \bar{t_2}, \ldots, \bar{t}_{|\bar{V}|} \right)$, then the information in (C) and (D) is sufficient for voter $i$ to make transfer delegation decisions that weakly increase the probability of a correct collective decision. For most voters, this information will be necessary as well as sufficient.}\\ 

\textbf{Proof.} Under assumption (C), the voting weight for voter $j$ in $\bar{V}$ is $\bar{t_j}.$ Denote the optimal weight for voter $j \in \bar{V}$ as $w_j$, consistent with our notation in the rest of the paper. Voter $i$, who is making the potential transfer delegation decision, is not a member of the set $\bar{V}$. Based on assumption (C), voter $i$ is able to discern the set of winning coalitions if voters are restricted to $\bar{V}$: A subset $S \subseteq \bar{V}$ is a winning coalition if and only if $\sum_{j \in S} \bar{t_j} > 0.5 \sum_{j \in \bar{V}} \bar{t_j}.$ I.e., winning coalitions consist of groups who have more than half of the voting rights.\\
\indent Parallel to the discussion preceding Lemma 3.5, define:
$$T^*_{-i} = \displaystyle\min_{A \subseteq \bar{V}, B = \bar{V}/A} \left| \sum_{j \in A} \bar{t_j} - \sum_{k \in B} \bar{t_k} \right|.$$
Because voter $i$ is minimally decisive, it must the case that $\bar{t_i} > T^*_{-i}$. By assumption (C), voter $i$ has enough information to determine the set of winning coalitions, and can observe the structural cases in which voter $i$ is potentially decisive. Assuming a rich enough environment to rule out ties, consider the winning coalition, $W_{m} \subseteq \bar{V}$ and the corresponding losing coalition, $L_{m} = \bar{V}/W_{m}$, that characterize the minimum difference, $T^*_{-i}$, and a small enough voting rights delegation $D_{small}$ by voter $i$ so that only outcomes involving this pair of coalitions are affected.\\
\indent If $W_{\bar{V}}$ is the set all winning coalitions considering only voters in $\bar{V}$, then the probability of making the correct collective choice is:  
$$P =  \sum_{S \in W_{\bar{V}}} \prod_{j \in S} p_j \prod_{k \notin S} (1-p_k).$$
Consider the term in this sum associated with $W_{m}$:
$$ T_{w} = \prod_{j \in W_{m}} p_j \prod_{k \in L_{m}} (1-p_k).$$
The key question is whether voter $i$ should delegate $D_{small}$ to one or more voters in $L_{m}$ in order to make $L_{m}$ a winning coalition with the consequence that the term in P changes from $T_{w}$ to:
$$ T_{l} = \prod_{j \in L_{m}} p_j \prod_{k \in W_{m}} (1-p_k).$$
In order to make this choice, by Lemma 4.2 the voter needs to answer the question:
$${\sum_{j \in W_{m}} w_j} \lessgtr {\sum_{k \in L_{m}} w_k}?$$
But then assumption (D) becomes necessary. To apply Lemma 4.2 to answer the question, voter $i$ needs to know the optimal weights for all the voters in $\bar{V}$. If the inequality is less than, then voter $i$ should delegate to one or more members of $L_{m}$, reversing the structure flowing from the voting rights profile $\bar{t}$ that made $W_{m}$ a winning coalition.\\
\indent If the inequality is greater than, voter $i$ could move to the winning coalition creating the next smallest gap between the voting rights in a winning and a losing coalition. Any transfer delegation made with respect to those two coalitions would also need to satisfy the constraint not to reverse the correct structure for the first set of coalitions, the coalition pair that created the smallest gap. Much more generally, voter $i$ could attempt to use an algorithm that considered all pairs of winning and losing coalitions with respect to which voter $i$ is decisive. It is likely that a maximizing algorithm is possible at least by brute force, but we leave that possibility to future work.\\
\indent If a voter has a low number of voting rights, then it is clear that the information embodied in assumptions (C) and (D) is necessary as well as sufficient. For example, if the voter is only decisive with respect to one pair of winning and losing coalitions, the exact structure becomes crucial. The voter cannot use some simple rule like delegating to the party with highest ratio $\frac{w_i}{\bar{t_i}}$. That might cause the losing coalition to win when it would be better if it remained the losing one in the pair. The optimal weights information becomes much more important relative to information about the profile of voting rights exercise for voters with very large shares of the total voting rights. For example, if a voter has almost all of the voting rights and knows the optimal weights for the rest of the voters, ignoring the profile of voting rights exercise will not matter much. The voter can delegate in a way that comes close to replicating the pattern of optimal weights. But for most voters, knowing both the profile of voting rights exercise and the optimal weights for the other voters will clearly be necessary as well as sufficient conditions. $\square$

This Proposition indicates the difficulties inherent in achieving optimal results using one-step transfer delegation. First, there is a coordination problem. The Proposition avoids this problem through assumption (B). In an actual one-step transfer delegation, delegation is occurring simultaneously and no one delegate will know the result of the delegation choices of all others prior to making their own delegation decision. On the information side, under assumptions (C) and (D), the voter needs to know $\bar{V}$, the set of participants who will actually exercise their voting rights, the number of voting rights each participant will exercise, and the optimal weight for every voter in $\bar{V}$. The last condition requires weight omniscience with respect to the set $\bar{V}$. Furthermore, these information requirements will be necessary for all or almost all participants in order for them to delegate by explicit transfers optimally.

One hope is that multi-step transfer delegation may make the challenge of optimal transfer delegation less difficult by making coordination easier. In a multi-step transfer delegation, it is plausible to assume that all voters can observe the outcome of the previous step. However, if there are a fixed number of steps, the exact same coordination problem will tend to arise as in one-step transfer delegation. Voters must execute one-step transfer delegation on the final step.\footnote{We say ``tend" because there are some circumstances involving voter knowledge of the epistemic competencies of others combined with particular social network properties that make coordination to achieve an optimal outcome for multi-step delegation possible. But the required assumptions are very strong. See Proposition 5.3 on page \pageref{connected_star} and the accompanying discussion infra.} We discuss transfer delegation that admits an unlimited number of steps next. That mode of transfer delegation also involves information needs, potentially as pervasive as those delineated in Proposition 5.1. We will examine some of the information assumptions that appear in the relevant body of literature and how those assumptions affect epistemic outcomes. 

One of the assumptions that appears in several of the papers we discuss in the next subsection we call \emph{high local delegative competence}: Voters have high delegative competence with respect to a limited number of other voters even though they have little knowledge of the epistemic competencies of the vast majority of other voters. This assumption suggests a general point that provides important perspective with respect to the results in Proposition 5.1. When governance involves a small group of individuals who know each other well and engage in repeated collective group decisionmaking, voters may well know the epistemic competencies of others and how delegation might affect the probability of making the correct collective decision. Assuming a shared epistemic objective and that all voters know both the views of every other voter through deliberation prior to a vote or otherwise and every other voter's optimal weight, there is no reason even to delegate. Each voter can use the Optimal Weighting Theorem to compute the decision that will maximize the probability of making a correct group choice and then vote accordingly. The result will be a unanimous vote for the more promising of the two alternatives.

In the more general case involving a large group and a setting in which voters only have knowledge about a small subset of other voters, there is no immediate epistemic solution. Nonetheless, the fact that there is a degree of local delegative competence has the potential to strengthen the epistemic properties of multi-step delegation. It is not a surprise that scholars have considered the impact that might accrue by making the assumption of high local delegative competence. In the next section, we discuss results from the literature and extend them by discussing voter behavior and coordination methods than strengthen the epistemic properties of liquid delegation.

\subsection{Transfer Delegation with Unlimited Steps: Liquid Democracy}\label{liquid}

One form of transfer delegation of both theoretical and practical interest for DAO governance is liquid democracy. Advocates such as \citet{Hardt_2015} state the case for liquid democracy versus direct democracy and representative democracy. Direct democracy is the method of many DAOs. Token holders, the population, vote directly on every proposal concerning the DAO. As \citet{Hardt_2015} note, although direct democracy offers "control, accountability and fairness," it is also the case that ``engagement falls off in large groups since voters generally do not have time or expertise to vote on everything," a phenomenon that is extensively present for DAOs.\footnote{See note 1 supra.} Representative democracy in which representatives are elected for fixed terms can enhance expertise but has other issues. \citet{Hardt_2015} note that these issues include ``problems with transparency, accountability, high barriers to entry in becoming a representative, abuses of power, focus on superficial aspects of candidates, and decisions being influenced by the election process itself, e.g. election cycle effects."

Liquid democracy allows voters to delegate and re-delegate their voting rights. Token holders can override their transfer delegations at any time to form new transfer delegations or to vote directly themselves. \citet{Hardt_2015} note that the ``ability to override and vote directly gives the control, fairness, and transparency of direct democracy" while at the same time securing the ability to be represented by others with more expertise.

An initial question is whether liquid democracy with no limit on the number of rounds will reach a stopping point. One problem is \emph{cycles}, which occur if the transfer delegation process returns the voting rights to the original delegator after a certain number of rounds. For example, in successive rounds A delegates to B, B delegates to C, and C delegates to A. Cycles have no stopping point. Even without cycling, there is the question of the existence of an equilibrium in which an allocation of voting rights is reached at which no voter wants to change their transfer delegation. And if there are equilibria, there is the further question of whether the transfer delegation dynamics inherent in a liquid democracy instance reach an equilibrium. Finally, even if an equilibrium is attainable, there is the epistemic question of whether it is a good outcome with respect to the probability of making a correct collective decision. 

Many results are quite negative. For instance, \citet{Escoffier_2019} examine implementation of liquid democracy in social networks. They start by eliminating the possibility of cycles by simply assuming that voters caught in a cycle will abstain. Even after taking that step, they find that the existence of a Nash-equilibrium is not guaranteed in general, and that it is  guaranteed independent of the profile of voter preferences if and only if the social network is a tree, that is, in addition to the absence of cycles, every voter is connected to every other voter by some path. But then they show that for some tree social networks and reasonable transfer delegation dynamics (such as changing transfer delegation at each round only if doing so improves the outcome for the delegator), the liquid democracy process does not converge. 

Putting aside the very real possibility that the liquid democracy process will not terminate if the number of steps is unlimited, both theoretical and empirical work indicate that from an epistemic standpoint there is a danger of \emph{over-delegation}, the concentration of excessive voting power in the hands of one or a few delegates. Consider the following example in which liquid democracy results in an extreme case of the whale problem discussed on page \pageref{whale_problem}:\\

\textbf{Example 5.2.} Assume the same framework as Example 2.1. There are 3000 voters with weight $w_i=0.1$ ($p_i \approxeq 0.525$) each with 1 voting right, and one voter with weight $w_i=1$ ($p \approxeq 0.731$) who has 10 voting rights. Majority rule based on voting rights is the decision criterion. Assuming independent competencies, the voting rights proportions exactly match the optimal weight proportions. If all holders fully exercise their voting rights, the probability that the collective decision will be correct is $0.997$, and the more competent voter has negligible effect. Now suppose that the voters use a liquid democracy approach to delegate before the vote, that each voter can ascertain the competence of any other voter, and that voters search exhaustively for a voter with higher competence and then delegate to that voter, only stopping the process if they can find no such voter. Then all voting rights will be delegated to the single high-competence voter, and the probability of a correct collective decision will plummet to $0.731$ from the majority vote outcome of $0.997$. Each low competence voter has thrown out their independent information, which would otherwise have made a correct collective choice more likely. $\square$\\

Unfortunately, the work of several researchers suggests that this result is not an anomaly, but may be in some sense be typical. We discuss four articles. Three of them are theoretical and use high local delegative competence assumptions, which we have mentioned increase the potential for liquid democracy to have good epistemic outcomes.

\citet{Kahng_2021} assume that each voter $i$ chooses from among a set of voters known to $i$ an unordered approved subset of voters $J$ such that $p_j-p_i > \alpha ~ \forall~j \in J$ where $\alpha > 0$ is a parameter common to all voters. They consider local transfer delegation mechanisms, which include delegating randomly to a member of the approved subset but do not include choosing the most competent member among them as a mechanism. Roughly speaking, they find that there are no local mechanisms that can both create the prospect of some level of positive gain and guarantee that losses are below any particular minimum as the number of voters gets large. They note the reason for the result: ``The main idea [underlying the result] is that liquid democracy can correlate the votes to the point where the mistakes of a few popular voters tip the scales in the wrong direction. As we show ... this is unavoidable under local transfer delegation mechanisms, which, intuitively, cannot identify situations in which certain voters amass a large number of votes."\footnote{\citet[p. 1238]{Kahng_2021}.}

In response to these results, \citet{Kahng_2021} develop a ``non-local mechanism" that does meet both properties, positive gain and minimal loss. A key feature of the algorithm is that it caps the number of votes that can accrue to each delegate. The cap is pretty severe, being $o\left(\sqrt{ln(n)}\right)$ where $n$ is the number of total votes. Even if $n$ is in the millions, each delegate is limited to a handful of votes. A likely pattern is a large number of delegates, each with a similar number of delegated votes. This pattern is reminiscent of many forms of representative democracy which have the property that voters are allocated in roughly equal numbers or at least numbers of the same order of magnitude across a fixed number of representatives. This pattern clearly blocks over-delegation. The analogy is not precise because voters in representative democracy do not have the ability to allocate their votes to representatives outside of their voting jurisdictions. But the analogy strongly suggests that it would not be a surprise if some forms of representative democracy proved superior to liquid democracy. Some of the results we discuss next suggest that direct democracy in the form of majority voting also may be superior to liquid democracy. Thus, contra the idealism that liquid democracy combines the best features of both representative democracy and direct democracy, it may be epistemically inferior to both of them individually.

\citet{Bloembergen_2019}, in a section of their article in which they adopt the \citet{Kahng_2021} framework, assume that voters choose the highest competence delegate (including self-delegation) and obtain similar results to the \citet{Kahng_2021} findings. As the network degree increases, the probability of liquid democracy reaching a correct majority outcome decreases along with a decrease in the number of ``gurus," those voters remaining after transfer delegation is completed. They conclude: ``Simply put, giving all the weight to a small group of gurus increases the chance of an incorrect majority vote, assuming that gurus have a less than perfect accuracy."\footnote{\citet[p. 1802]{Bloembergen_2019}.}

In an extensive and well-constructed empirical analysis of liquid democracy in practice spanning 250,000 votes and 1,700 proposals across 18 DAOs, \citet[p. 3]{Hall_2024} find that ``delegation is somewhat lumpy, leading some delegates to amass considerable voting power, consistent with theoretical concerns about over-delegation."

In an elegant paper, closely related to the work here, \citet{Mooers_2024} engage in a theoretical and experimental examination of liquid democracy, comparing it to majority voting and majority voting with abstention (``MVA").
The authors see themselves as ``studying variations of majority voting,"\footnote{\citet[p. 21]{Mooers_2024}} and perhaps for this reason, votes are indivisible. As a consequence, MVA consists of either voting or entirely abstaining, and transfer delegation also is all or nothing. 

The authors conduct two experiments. Both experiments and the associated theory are based on the assumption of independent competencies. In experiment 1, one or more experts all have the same probability, $p_i = p$, of arriving at the correct decision on their own, and this probability is publicly known. Non-experts have competencies drawn randomly from a distribution with support $[0.5, p]$. Non-experts know their own competencies, but this information is private to each non-expert. Delegations shift votes to a randomly selected expert. In the second experiment, competencies are not fixed or even known precisely. Participants engage in a perceptual game that is the competency task. Trial runs of the game provide the estimates of competencies used to select the top $20\%$ of participants as experts. 

In both experiments the probability of a correct collective choice is highest for majority voting, closely followed by MVA, with liquid democracy falling quite far behind, contrary to theoretical predictions based on calculating equilibria under the variants. These patterns result from over-delegation: ``[I]n both experiments, participants delegate with very high frequency, much more frequently than they abstain and two to three times more frequently than optimal in the experiment for which we have precise predictions."\footnote{\citet[p. 38]{Mooers_2024}} The authors carefully consider some possible reasons for this outcome but do not reach a firm conclusion. At the start of the paper, the authors note a ``fundamental problem ... even if the experts are correctly identified, delegation deprives the electorate of the richness of noisy but abundant information distributed among all voters."\footnote{\citet[p. 1]{Mooers_2024}} At the end of the paper, the authors suggest that MVA ended up being more robust than liquid democracy in their framework because abstention effectively delegates to all other voters while liquid democracy delegates only to the expert group.\footnote{\citet[p. 38]{Mooers_2024}} They conclude: ``Liquid democracy should be tested outside of the lab, but our tentative conclusion is that, on informational grounds alone, the arguments in favor of Liquid Democracy should be considered with caution."\footnote{Id.}

These problems and trade-offs are not inevitable but stem from the defects of the choice systems considered, especially when combined with assuming that votes are indivisible. The strength of partial abstention versus transfer delegation, majority rule, or MVA with indivisible votes is crystal clear in the framework of experiment 1 of \citet{Mooers_2024} and arguably also in the frameworks of \citet{Kahng_2021} and \citet{Bloembergen_2019}. All three frameworks assume high local delegative competence: Participants have strong information about both their own competencies and the competencies of at least a subset of other voters.

Consider first experiment 1 from \citet{Mooers_2024}. Each non-expert voter knows their own epistemic competence, $p_i$, and the highest level of competence among all voters, the expert competence, $p$. Each voter has one voting right. Using the terminology and results from Corollaries 3.2, 3.3, and 3.4, all voters know $R = \frac{ln(p)}{ln(1-p)} = w_p$ where $w_p$ is the expert voters optimal weight, and non-expert voter $i$ can abstain on all but $t_{i}^{*} =  \frac{w_i}{w_p}$ of $i$'s unit voting right. If all voters proceed in this manner, the voting weights will be proportional to the optimal weights under the Optimal Weighting Theorem, and the probability of the correct collective decision will be maximal. There are no trade-offs. Each non-expert voter's independent information, however modest, will enter into the social decision appropriately weighted.

\citet{Kahng_2021} and \citet{Bloembergen_2019} assume that voters can do close comparisons of their own epistemic competencies with at least a subset of other voters. In \citet{Kahng_2021}, voters are able to identify a subset whose competencies exceed the voter's competence by at least some number $\alpha$. In \citet{Bloembergen_2019}, voters are able to rank members of the subset and the voter with respect to competence. These frameworks strongly suggest that each voter $i$ knows their epistemic competence, $p_i$. What the voters do not know is the highest competence level among all voters. However, each voter has a single voting right. The version of partial abstention exposited in Example 3.7 can achieve the highest possible probability of a correct collective choice to any degree of approximation desired.

Unlike partial abstention, majority voting, MVA, and transfer delegation with indivisible votes all result in sub-optimal voting weights and a lower probability of a correct collective choice in frameworks where voters know their own epistemic competence. If we permit divisible votes in the instance of transfer delegation (regardless of the number of steps), achieving an optimal result might be possible, but, as indicated by Proposition 5.1 in the case of one-shot transfer delegation, the coordination problem is much more difficult than the simple approaches for partial abstention that follow from Corollaries 3.2, 3.3, and 3.4 and from Example 3.7. There may be reasons to use methods other than partial abstention in some framework, but with independent competencies and voter knowledge of their own competencies, there is not.\footnote{This conclusion becomes less clear if voters do not have exact knowledge concerning their own competencies.  Consider, for example, experiment 2 in \citet{Mooers_2024}. In that experiment, voters do not have precise knowledge of their own competencies or the competencies of others. Uncertainty about the competency of others is not a problem in the partial abstention approach because approximation as described in Example 3.7 can be used in the absence of exact knowledge about the competence of the most adept voter. On the other hand, uncertainty about own competencies adds complexity. For example, if voters differ in the precision of own-competency knowledge, it may be optimal to adjust the weights accordingly, especially if there is social risk aversion.} Based on the literature discussed in this subsection, liquid democracy appears to be a particularly frail competitor.

To emphasize the relative strength of partial abstention versus transfer delegation, we consider and analyze transfer delegation in a way that puts it in a more favorable light. One striking element of some of the current models of transfer delegation is that they seem to assume sub-optimal behavior by delegating voters. In these models, voters know the competencies of a subset of other voters, but they typically delegate to only one of them. As discussed in subsection \ref{list_based}, in the case of independent competencies, the optimal single-step strategy would be to delegate among the voters in the subset using the optimal weights specified in the Optimal Weighting Theorem. As we prove subsequently, in some environments this approach also works well when there are multiple steps.

Consider a directed graph description of a social network in which the nodes are voters, and each voter knows their own epistemic competence (Assumption 3) and the epistemic competencies of some subset of the other voters. Formally, for each voter $i$, there is a subset $N_i \subseteq N$ of the voters whose epistemic competencies are known to $i$. By Assumption 3, $i \in N_i$, and, thus, $N_i$ is non-empty. Define a \emph{star cluster} as a directed subgraph consisting of edges pointing from $i$ to each member of $N_i$. Note that if $N_i = \{i\}$, then the directed subgraph only contains a directed loop from $i$ back to $i$. The \emph{directed star cluster graph}, $G$, consists of the union of all the star clusters, one for each member of N. Star clusters are \emph{overlapping} if they share at least one node. A \emph{path} between two nodes exists if there is a sequence of overlapping star clusters that includes both nodes. Define a directed star cluster graph, $G$, to be \emph{connected} if there is a path between every pair of nodes in $G$. Then we have the following result:\\

\textbf{Proposition 5.3.} \emph{Assume independent competencies, a shared epistemic objective, cognizable probabilities, the ability of any voter to observe the outcome of previous transfer delegation rounds before delegating in the next one, and voting rights that can be divided into arbitrarily small units. Suppose that the knowledge of voters about the epistemic competencies of other voters is characterized by a directed star cluster graph, $G$. If $G$ is connected, then using transfer delegation, voters can coordinate on an equilibrium that results in the highest possible probability of making a correct collective decision in at most three steps.}\label{connected_star}\\ 

\textbf{Proof.} We show there is a three-step coordination method that results in maximizing the probability of making a correct collective decision. In a first step, each voter $i$ delegates to members of $N_i$ based on their optimal weights, which are known to $i$:

\begin{quote}
\textbf{Step 1.} Each voter $i$ delegates all $t_i$ voting rights held by that voter proportionally to all members of $N_i$ according to the optimal weights from the Optimal Weighting Theorem defined in equation (\ref{optimal_weights}) on page \pageref{optimal_weights}.
\end{quote}

Claim: After this step is complete, all voters have weight omniscience. To demonstrate the truth of this claim, start with some voter $i$. For every voter $k \in N_i$, the star cluster emanating from $i$, voter $i$ knows $p_k$ and thus the optimal weight, $w_k$, for voter $k$. Because $G$ is connected, the cluster $N_i$ overlaps with at least one other cluster. If instead, $N_i$ were isolated, there would be no path from $i$ to voters outside of the cluster. Suppose $N_i$ overlaps with $N_j$. Then voter $i$ knows the epistemic competency and optimal weight for at least one voter $k \in N_j$. But voter $i$ also knows the relative weights of all voters in $N_j$ because those weights are evident from voter $j$'s delegation among those voters in step 1. Therefore, voter $i$ can compute the optimal weight, $w_y$, of every voter $y \in N_j$ based on knowing $w_k$ and the ratio $\frac{w_y}{w_k}$, which is equal to the ratio of voting rights that voter $j$ made by delegation between voters $k$ and $y$ in step 1. For any $x \in N$, voter $i$ can apply this approach recursively along the sequence of clusters for a path from $i$ to $x$ in order to learn $w_x$, and the claim is established.\\
Given that all voters have weight omniscience, there are multiple ways to coordinate so that the probability of making the correct collective choice is as high as possible. One simple way is use the following second step:

\begin{quote}
\textbf{Step 2.} Each voter $i$ delegates their voting rights to all voters, including $i$ by the self-delegation step of retaining and exercising some voting rights, in proportion to the optimal weights of the voters.
\end{quote}

Then a third step:

\begin{quote}
\textbf{Step 3.} Each voter exercises all the voting rights held by that voter after step 2.
\end{quote}

Because each voter delegated using proportions based on the optimal weights, the aggregate allocation of voting rights after step 2 will be proportional to the optimal weights. If all voters exercise all of their voting rights in step 3, then the probability of making the correct collective decision will be as high as possible. $\square$\\

Proposition 5.3 indicates that given the information structure in which at least some voters know the epistemic competencies of a subset of other voters, it is possible in some instances to maximize the probability of making a corrective collective choice using transfer delegation. Over-delegation effects are absent along with the consequent possible degradation of decision probity.  We have only demonstrated this result for the case of social networks that are connected directed star cluster graphs.\footnote{If we had a directed star cluster graph that is disconnected, we could reach a solution if at least one voter in each star cluster reported their epistemic competence, knowing that these reports are truthful given the assumption of a shared epistemic objective. But then we are not relying on transfer delegation itself as in the proof above in which each voter $i$ relies only on that voter's knowledge about the epistemic competencies of voters in $N_i$, the star cluster associated with $i$, along with knowledge of the observed delegations from the first round of delegation. As discussed in note \ref{candidates_only} and the accompanying text on page \pageref{candidates_only} supra, the goal is to compare the effectiveness of ``candidate" approaches such as transfer delegation and partial abstention.}

The relative strength of partial abstention remains. It will maximize the probability of a correct collective choice even if the social network is a directed star cluster graph but is not connected. Partial abstention does not require any knowledge of the epistemic competencies of others. It would work in the degenerate case in which each star cluster is a singleton and there is no path between any two voters.

\subsection{Sortition}

Use of sortition in social choice systems is a rich topic accompanied by a large literature. Full consideration of the epistemic aspects would be lengthy. We limit ourselves to a few basic observations.

Classic sortition consists of choosing a random sample of potential voters as a decision group on behalf of the entire collective. There are lots of advantages. Consensus mechanisms such as Algorand choose a new sample for each governance step, a move that means the selected group cannot be targeted for bribery or undue influence because their voting power is one and done.\footnote{The sortition mechanism employed by Algorand and by other proof of stake consensus mechanisms such as Ethereum have been very successful. There is a shared epistemic objective of repeatedly achieving an honest consensus for blocks and very strong economic incentives for the voters, the network nodes that choose to participate, to adhere to this objective. It is easy for nodes to get up to speed and operate effectively because they are engaging in the same task over and over again. The general context we are considering does not have these features. Decisions generally are ones of first impression and voters do not opt-in subject to powerful rewards and penalties that motivate reaching the correct result.} The same cannot be said of representatives who are elected for fixed terms of office. Using a sample of voters may also make deliberation easier and investment in becoming more informed more attractive because the decision group is smaller and individual members are more likely to be decisive.

From an epistemic standpoint, there is a tradeoff if independent information is widespread. The information in the hands of excluded voters does not enter into the decision process. In the case of independent voters with equal epistemic competencies and a one-person-one-vote approach, Condorcet's Jury Theorem runs in reverse because reducing the number of participating voters lowers the probability of making a correct collective choice, and a drastic reduction in the total number may have a very big impact.  The widespread information case may be quite common. For instance if voters also use the services of a DAO, each voter may have a unique perspective that adds some value, however small, to assessing future approaches. In that case, every voter is receiving at least one independent signal not received by any other voter.

If voting using optimal weights is the method, sortition reduces the probability of making a correct collective decision in the independent competencies case if voters who have positive weights are excluded. The situation is as if we had the full population participating but incorrectly set these voters' weights to zero. By the Optimal Weighting Theorem, the probability of a correct collective decision will fall as a result of this miss-weighting.

One well-known problem of sortition, discussed, for example, in \citet{Lever_2023} is that differential willingness to serve may create distortions in the representativeness of the selected group. The actual group consists of those randomly selected minus the unwilling. The epistemic version of this problem is that the unwilling may have particular characteristics that reduce the epistemic outcome if they exclude themselves. For example, if potential participants with higher expertise are less willing to serve due to the time commitments to deliberations and voting processes required but would be perfectly happy to be informed voters otherwise, sortition may result in a group with lower average epistemic competence than the population. In the case of independent competencies, the selected group will have a lower probability of reaching a correct collective choice in any system with fixed weights such as one-person-one-vote as well as any system that employs optimal weights.\footnote{With respect to the fixed weights case, see Lemma 3.5 supra.}

The random nature of sortition also creates uncertainty, especially if the chosen group is small. By chance the group may include voters with below average competencies when competencies are independent, and when they are dependent, the group may contain voters that have received similar independent signals to the exclusion of others. On the other hand, in a flooding situation, the selected group may by chance include a disproportionate number of highly competent voters whose influence might otherwise be washed out in a system that does not use optimal weights in the voting rule.

Overall the epistemic consequences of sortition are complex and may depend on particular group draws. But it is clear that there is potential for a major epistemic impact compared to any particular baseline voting rule that is the alternative for choice by the whole population. 

\subsection{The Delegate Slate}\label{slate}

So far, we have considered the situation of a fixed delegate slate: The set of possible delegates is fixed. But there is another aspect of transfer delegation that often will have a major impact: adding and subtracting  delegates who are not otherwise in the system as voters. 

Consider the independent competencies setting in which the voting rule employs optimal weights across all voters including delegates. Proposition 4.1 indicates that adding delegates with epistemic competencies greater than $0.5$ weakly increases the probability of making a correct collective decision. 

The same is true in the context of majority voting (one-person-one-vote) and independent competencies if we add delegates with high enough competencies relative to the existing group. We have the sufficient condition of \citet{Ben_Yashar_2017} discussed on page \pageref{pair} which is $w_i + w_j > w_1$ where $i$ and $j$ are the added voters and voter $1$ is the most competent voter in the existing group.

For rules other than the use of optimal weights, it is clear that all of the dangers identified in section \ref{participation} can arise if delegates are added or subtracted from the slate: the flooding danger, the dependency danger, and the epistemic danger from whales. The exact outcomes mix considerations of adding participants and how the transfer delegation will work. For example, when many low competence delegates are added, the flooding danger will occur only if many of these delegates receive actual delegated voting rights. 

Although it is hard to say anything in general, it is clear that delegate slate choices may have a big impact. Policies such as creating very visible menus of delegates to draw extra delegative participation from relatively passive voters must be done with care. The composition of the menu and whether there are limits on transfer delegation to each delegate may make a big difference. In subsection \ref{shades} we discuss some possible attempts to use delegate slate formation to improve the epistemic performance of DAOs. 

\section{Shades of Decentralization and Epistemic Supplements} \label{degrees}

It is clear that DAO governance may struggle with epistemic performance, the ability to reach an acceptable probability of making a correct collective choice or at least one that does well in that regard with respect to the resources available. There are some potentially effective responses that may be considered compromises versus idealized versions of decentralization such as governance by majority vote in an environment in which voting power is not concentrated structurally or in practice. If such ideal forms of DAO governance are not attainable in practice or if they have significant negative epistemic consequences, then that degree of decentralization is not viable.  We then need to consider shades of decentralization that have decentralized features but depart from idealizations that we may wish worked on their own. We briefly discuss four examples in this section, the last three of which rely on markets to aggregate information for purposes of governance.

\subsection{Systematic Independent Transfer Delegation by Whales}\label{shades}

Proposition 4.1 indicates that DAO participation can be enhanced with no epistemic cost in the optimal epistemic environment. This environment requires that voting conform to the optimal weights specified by the Optimal Weighting Theorem. That is a tall order for a decentralized system. But something close may be possible if there are entities or individuals who hold a substantial portion of DAO ownership, perhaps 20\% or more. Suppose there is at least one such ``whale." Normally, whales are seen as a threat to decentralization because large stakes can carry with them explicit or implicit control of the DAO.

Suppose, however, that the whale uses its stake to create a collection of carefully selected delegates, each of whom are granted independence from the whale with respect to voting. The whale can select delegates, some of whom have no current connection with the DAO, who bring diverse, valuable viewpoints to governance and then weight them by transfer delegation amounts in a way that approximates optimal voting weights. Choosing and motivating delegates to perform actively and continuously in this framework creates a powerful block of votes that is likely by itself to exhibit excellent epistemic performance. Although operating this kind of ``governance machine" does not create precisely the optimal epistemic environment, it makes added participation by others less likely to have a negative epistemic impact on outcomes. 

The whale's large stake creates a substantial economic incentive to do an excellent job of structuring and maintaining such a governance machine. For instance, a whale with a 40\% interest will reap 40\% of the economic returns from good governance. This potential return can justify investment of substantial resources in terms of money and time. The same cannot be said for diverse small holders each acting on their own because they each would reap only a small portion of the returns of any such investment.\footnote{One intriguing possibility is for a more collective implementation of systematic independent transfer delegation. For example, instead of implementation by a whale, a subgroup of smaller holders or the entire group of all DAO token holders could elect a delegate management entity of some kind to carry out systematic independent transfer delegation. The subgroup or group could delegate their voting rights to the delegate management entity as a first step. The trouble with this approach is that it simply shifts the primary governance problem we are addressing back a step. How do we know that the subgroup or group will be good at choosing the delegate management entity or at monitoring that entity's performance? The Jensen-Meckling problem is present: Small stakeholders individually may be rationally passive to a large degree with respect to the selection and monitoring functions because of their modest economic stakes. The benefits of their efforts would flow overwhelming to others, creating the serious collective action problem that Jensen and Meckling identified.} Even in the case of a 40\% whale, the owners of the other 60\% free ride on the whales's efforts. This dilemma is the Jensen-Meckling problem, first described by \citet{Jensen_1976}, who point out that the problem only disappears when ownership reaches 100\%. In that case, the owner is motivated to make the fullest possible effort that is economically justified. Owners with less than 100\% are likely to exert lower levels of effort, causing the result to fall short of its potential. The contestable control alternative presented in subsection \ref{contestable} overcomes this problem.

There are examples that appear have some characteristics of systematic independent delegation by whales. \citet{Amico_2021} describes a16z's token delegate program, which includes transfer delegation of a large portion of a16z's substantial token holdings in several DAO projects to parties such as non-profit organizations, startups, university organizations, and crypto community leaders. Some of these parties are not token holders in the DAOs, and none of them is required to be such. a16z aims at a ``diversity of perspectives" in the mix of its delegates as well as ``active users who understand the protocol and want it to grow, but otherwise lack sufficient voting rights to participate meaningfully." This language sounds like an effort to select epistemically independent delegates each of which have high epistemic competence, a strong formula for epistemic performance. Also, a16z, as a leading venture capital firm in the cryptocurrency space, is likely to have high delegative competence.

Delegation is a key function of management, whether centralized or decentralized. And for independent transfer delegation by whales to be properly incentivized, the DAO must be an economic DAO, which we defined in subsection \ref{epistemic} as one for which the aggregate token value reflects the value of the project. Then a whale's large token share creates the proper incentives for the whale to delegate effectively and to identify and introduce new delegates who will add value. In the background is the need for DAOs to compete with organizations that employ centralized management.

How much does a governance machine created through independent transfer delegation by whales shade decentralization? Much depends on whether the delegates are truly independent from the whales. It is possible to create a contract to embody the independence with a long enough term so that it is meaningful, but there are other factors, such as a desire to earn a continuation, that may bias delegate votes.
There also are regulatory factors. Some approaches such as the FIT21 legislation passed by the U.S. House of Representatives and pending in the U.S. Senate create quantitative ownership tests that determine regulatory treatment.\footnote{H.R. 4763 is known as the Financial Innovation and Technology for the 21st Century Act (``FIT21"). H.R. 4763 was approved by the House Financial Services Committee on July 26, 2023 and passed by a strong bipartisan majority of the full House on May 22, 2024, creating the possibility of enactment if approved by the Senate and signed by the President. \citet{HR4763}; \citet{HR4763pass}.}  In the case of FIT21, 20\% ownership or more by one token holder is disqualifying for the most favorable regulatory status. A good amendment would be to disregard truly independent transfer delegations for sufficient time periods and in small enough pieces. That would permit whales with strong incentives to delegate well to be part of the framework without triggering adverse regulatory consequences.

\subsection{Futarchy}

Futarchy is the use of prediction markets to make decisions. \citet{Hanson_2013} provides a good summary of the basic ideas. Use of futarchy in DAO governance is of great current interest, and projects such as \citet{Butter_2025} have created implementations.

Futarchy works by specifying an issue, creating a prediction market, and then taking action based on the market outcome. For example, the proposition might be something like: If the DAO implements policy X, decentralized exchange turnover will increase to at least Y billion by January 1, 2026. A market for the proposition is created which pays \$1 per contract on January 1, 2026 if the proposition proves to be true and \$0 if it proves to be false. At some point, the DAO observes whether the market favors the proposition and acts accordingly. If the market favors the proposition, the DAO implements policy X, and the yes market remains open with the payoff on January 1, 2026 being \$0 or \$1 depending on the outcome. If the market does not favor the policy, the market is shut down because the hypothesized event, an outcome conditional on policy X being implemented, can never occur because X was not implemented.

As detailed in \citet{Snowberg_2013}, prediction markets have a good track record for accuracy. DAOs can use them instead of votes to decide certain issues.\footnote{It may be possible to go beyond merely replacing voting approaches with prediction markets. \citet{Airiau_2024} show that under certain restrictive conditions, including very particular utility functions, prediction markets can produce outcomes that approximate voting with optimal weights. We discuss approaches motivated by this result in the next subsection.} Creating a market is costly, and it will not work unless there is enough participation. Framing the subject proposition also is critical and requires both a choice of objectives and selection of measurable outcomes that embody the objectives. As a result, there are governance decisions surrounding both the choice to use a prediction market for a particular issue and how to implement the market in that case. Nonetheless, decision quality may be enhanced compared to voting, especially given some of the potential defects of voting systems that we have discussed. Futarchy is an alternative epistemic approach, one that depends on markets instead of voting to aggregate information. 

How much does the use of futarchy shade decentralization? Voting is being superseded, and the exact shape and embedded values in the use of futarchy for a particular decision might be determined by a handful of parties even if approved in a routine DAO vote. Nonetheless, market aggregation of information through a prediction market might be much more effective for some issues than the aggregation that occurs in voting. And DAO voters might prefer a prediction market approach for particular decisions. DAOs do delegate certain management functions to third-parties by contracting with them. Use of futarchy is at least somewhat analogous.

\subsection{Condorcet AI Agents}

\citet{Airiau_2024} in a paper titled ``Condorcet Markets" show that if traders are endowed with a particular logarithmic utility function of the price in a prediction market and the amount invested in that market, the utility being parameterized by single number $k$, then as $k$ tends toward infinity, the decision implicit in the prediction market price converges to what the decision under voting based on optimal weights would be.\footnote{\citet[Theorem 5, p. 515]{Airiau_2024}.} In the author's words: ``[T]his ... result shows that elections that are perfect from a truth-tracking perspective [ones that use optimal voting weights] can be implemented increasingly faithfully ... as the parameter $k$ ... grows larger."\footnote{\citet[p. 515]{Airiau_2024}.} Investment strategies in markets accomplish the otherwise difficult task of ``elicit[ing] truthful weights from agents."\footnote{Id. at 517.}

\citet[p. 517]{Airiau_2024} state that whether the approach can prove valuable in practice requires more research, and they note that the results are subject to ``at least four main limitations, which include use of standard jury theorem assumptions, the focus on only a one-shot interaction instead of allowing for iterations, the assumption that agents are price takers, and the fact that the study is based on ``very stylized utility functions that are the same for all agents." The standard jury theorem assumptions correspond to the structure here in the independent competencies case, that structure including elements such as a binary decision with a correct alternative. 

Three of these limitations disappear if we consider creating appropriate AI agents. These agents can be endowed with the precise utility function associated with approximating a ``perfect" voting system result based on optimal weights. We can specify a sufficiently large value of $k$ for the agents. We can require them to act as price takers and behave as if they are engaged in a one-shot interaction at each decision point. 

The assumption of independent competencies remains a barrier, but in some applications we may be able to do well enough by keying particular agents to particular information sources or by building in an approximation based on the independent signals framework discussed in section \ref{abstention} that included the possible use of discounts based on the number of agents that share sources. The agents might be able to compute this number with some degree of accuracy because elements such as the number of agents are observable aspects of the environment. And we have all the tools of AI, both known and not yet developed, to allow agents to learn and possibly coordinate through computation and communication. 

There is a possibility that we could build something analogous to systematic independent transfer delegation by whales, a block of AI agents that come close to operating among themselves using optimal voting weights, to provide an anchor or supplement to a DAO voting system.

In honor of \citet{Airiau_2024}, we might call agents in such a system ``Condorcet agents" or ``Condorcet AI agents." Creation of such a system is an exercise in imagination at this point. Much depends on questions such as whether AI agent information collection and processing could be strong enough to make a serious contribution to collective decisions and whether such agents could accurately estimate the quality of the information they collect in terms of epistemic competence. Nonetheless, it is clear that there is a  possibility of approximating a nearly ``perfect" voting approach using prediction markets. And it might work for at least certain applications. 
 
\subsection{Contestable Control}\label{contestable}

\citet{Strnad_2025cc} proposes a contestable control approach for DAO governance. This approach allows any party to initiate an auction to take temporary control of the DAO, itself contestable through a subsequent auction. Bidders promise to increase the token value by a specified amount and claim some portion of the resulting market capitalization increase. The bid that promises the largest value increment to the other token holders wins. Bidders submit deposits that guarantee performance and that will be slashed with compensatory payments to the other token holders if the DAO declines in value during the control period. As a result, the other token holders are guaranteed the promised increase in token value or its cash equivalent and are protected against losses by the deposits.

The operation of the auction mechanism exposes the control party to 100\% of the gains and losses in value of the entire DAO. As a result, the control party is in the same position with respect to incentives as a 100\% owner, and the Jensen-Meckling problem disappears. 

The auction mechanism is designed so that the bidder with the best business plan tends to win the auction. This quality is the epistemic aspect of the mechanism. As in the case of futarchy, the mechanism relies on markets rather than voting for epistemic performance.\footnote{What the auction mechanism adds that futarchy may not provide very well is properly incentivizing certain major innovations. Bidder business plans are private information, and the mechanism allows bidders to claim a bidder-specified portion of the surplus that the bidder's business plan will produce. Bidders are motivated to claim only what they need to cover their costs and make a normal profit because the higher their surplus claim, the weaker their bid is: The auction winner is determined by the bidder who promises to deliver the most surplus to the other token holders.\\
\indent If bidders fully revealed their business plans, which would be required for prediction markets to compare them intelligently, they would be in danger of not earning the surplus required to motivate creation of the business plan in the first place. After the business plan is made public, it can be applied without rewarding the inventor. The bidder may not have or be able to accumulate a large enough stake in the DAO prior to revealing the business plan to earn sufficient surplus given that token purchases reveal information and move prices upward and also that in some countries even moderate ownership stakes must be revealed. This difficulty is a version of what sometimes is known as the ``Grossman-Hart free-rider problem." \citet[pp. 13-14]{Strnad_2024}. The auction mechanism eliminates this problem entirely by allowing bidders to claim a specified amount of surplus without making any token market purchases prior to the auction.}

After the control period ends, the DAO reverts to its baseline governance state, typically a voting system such as the ones we have examined in this paper. As a result, the auction mechanism can provide guardrails for that voting system. If some of the problems detailed in the rest of this article threaten the performance of the DAO and there is an accompanying drop in token value, there are strong incentives for a party to step up and initiate an auction. The prevailing control party can reset or reform the voting system and then allow the DAO to revert to its baseline governance state.

How much does adding a contestable control mechanism shade decentralization? On the one hand, the mechanism temporarily centralizes control of the DAO. On the other hand, one purpose of the mechanism is to defeat the explicit or implicit entrenchment of control that seems to have characterized actual DAOs.\footnote{See note 1 supra.} In the face of a possible auction, neither whales or founders can secure on-going control. This feature allows revision when these traditional control parties are standing in the way of better outcomes for the DAO. For instance, if a whale implements an independent transfer delegation system for which there is substantial room for improvement, a bidder can gain temporary control and reform the system. In addition, as mentioned, the contestable control mechanism may create conditions that allow decentralized governance to function well by creating guardrails for desirable governance features that otherwise might be too risky. Although the mechanism involves temporary centralization, it may prove highly valuable or even essential for implementing successful decentralization. 

The contestable control approach only works well for economic DAOs, ones for which the token value market capitalization embodies the value of the project. In that case, the auction creates incentives for substantially improving project performance. As noted in subsection \ref{epistemic}, economic DAOs are not limited to DAOs that are commercial. 

For economic DAOs, there is an obvious governance objective that emerges from capital market pressures: maximizing market capitalization. For any governance decision, the right answer is the one that maximizes the market value of the project, an aspect that puts us in an epistemic world in which Assumption 1, that decisions have a single correct answer, is appropriate.\footnote{One concern is that this ``right answer" may embody shifting the DAO toward a web2 type of operation, leveraging network economies of scale to gain power over consumers and content creators. However, this danger is present already if the DAO token is traded because parties can gain control by buying up tokens. The only way that web3 can thrive in a capital market setting is if there is a large enough base of token investors that some web3 token projects will have market values that depend on promoting web3 characteristics.}  And it does so even in the face of the fact that some of the elements that drive market value themselves do not easily lend themselves to a right answer and are not commercial: various normative, political, or moral elements such as the value of participation in the baseline DAO decision processes. If these elements matter to participants, then embodying them in the DAO despite consequent, but lesser, sacrifices in operational efficiency will be reflected in its market value, and winning bidders in a contestable control contest will be the ones who can best implement them. Futarchy shares some of these characteristics, especially if the proposition that is the focus includes a prediction of the impact of a particular policy approach on the DAO's market value.

\section{Conclusion}

The major question that we started with is whether governance can be decentralized and at the same time be efficient in the sense of arriving at good policies and solutions. To address that question, we created two epistemic tests, one for the case where voters' judgments are independent and the other for the case where there are dependencies. These tests address situations in which there is a correct but unknown choice between two alternatives. The metric is the probability of arriving at the correct collective decision. 

We make some strong assumptions about participants' knowledge and motivations. Participants know their own epistemic competencies, the probabilities that they would make the correct decision on their own. Participants share the epistemic objective of achieving the correct collective outcome. In the case of dependencies, the total information set can be decomposed into a canonical list of independent signals, and each participant knows the probability that each signal received by the participant will result in the correct collective decision.  

Despite the strength of these assumptions and the limitation to cases in which there is a correct answer, we would want DAO governance to perform well in this environment, especially in the face of competing centralized alternatives. Some important questions considered by DAOs arguably do have a correct answer, and addressing them poorly may imperil the entire DAO enterprise whatever its broader aspirations might be. Good performance as well as decentralization are required for web3 ventures to succeed versus web2 competitors.\footnote{The line between questions that have a correct answer and ones that do not is not well defined in terms of choosing governance structures. Outside of an epistemic framework, the question becomes: What is the metric we will use to judge governance systems? If the metric is clear enough, we are back into an epistemic framework in which the goal is to maximize the probability of strong performance according to the metric. If there is no such metric, even approximately, then it is unclear that we can make systemic judgments about governance at all.} In addition, the outcomes of the epistemic tests have broader implications for governance.

We have shown that even with the strong assumptions we make, it is very unclear that methods such as various forms of delegation will lead to good results. The strongest candidate which emerges is partial abstention. In the independent competencies case, a partial abstention approach makes it possible for a decentralized governance system to maximize the probability of making correct collective decisions or come arbitrarily close to being able to do so. The degree of coordination required is very low, basically common knowledge of a particular reference number. In the case of competence dependencies, however, more is required. Assuming that the decomposition into independent signals is possible and that voters can identify them, voters must know how many other participating voters have received each signal that the voter receives in order to coordinate effectively to maximize the probability of making a correct collective decision.

Does partial abstention have a chance of being an effective governance method in practice? Many of the assumptions may fail, and perhaps dramatically so, in real world environments. Voters may not have very sound knowledge of their own competencies. Some may even have negative competencies, $p_i < 0.5$, and not be aware of it. The shared epistemic objective assumption will not be true if there are malicious actors. In partial abstention approaches, voters may abstain on a large number of voting rights, which creates an opening for malice because fewer voting rights are required to succeed at any particular attempted manipulation. This problem might be ameliorated by adding a final step in which abstained voting rights are delegated pro rata by the voting mechanism on the basis of the exercised voting rights. Adding that final step reduces the scope for malice, but the possibility of malice remains. Aside from actual malice, there is the possibility that voters will ignore the epistemic objective and simply use all of their voting power to promote their favored alternative. That approach undermines the epistemic qualities of the decision process because such voters will be over-weighted compared to what is optimal.

To be fair, all of these criticisms apply in one form or another to any of the decision approaches we have examined. But the questions remain: Could partial abstention work well enough in an approximate way? Is it plausible that a sufficient number of voters will share the epistemic objective of maximizing the probability of making a correct collective decision? If so, will voters be sufficiently adept at approximating their own epistemic competence? 

Then there are the difficulties of addressing dependencies. We have modeled dependencies as a decomposition into a set of independent signals. There are many other ways to do it. For example, we could have focussed on a covariance matrix. A major reason to consider the independent signals approach is that voters might be able to approximate optimal strategies when using partial abstention. As we have seen, one such optimal strategy is for each voter to discount signals received by dividing the signal weight by the number of other voters who receive the same signal. It is possible that voters may have an accurate sense that some of their information sources are widely shared and could guess at the appropriate discount for such signals.

The bottom line is that the real world case for using a partial abstention approach is unclear. Experimentation at the laboratory level and in functioning applications is required to study the effectiveness of the approach.

Some of our results with respect to direct participation have potential significance. Many see value in direct participation even if there are negative epistemic consequences. We identified an environment in which there is no such clash. In a voting system that employs optimal weights, added direct participation never hurts epistemic performance and is likely to have a positive impact. Departures from that optimal epistemic environment, however, make added direct participation of certain kinds a substantial epistemic threat. The system may be flooded with low competence or highly correlated voters, which can cause significant epistemic degradation. Voting rights whales, either large token holders or major delegates, may have way too much voting weight, which can have serious negative consequences.

Transfer delegation, including liquid democracy, does not fare any better. Epistemically appropriate transfer delegation requires overcoming difficult coordination problems and requires strong voter information about the epistemic and delegative competencies of other voters. In the epistemic environments we have studied, transfer delegation is dominated by partial abstention. Because partial abstention does not involve external relative weight intervention, knowledge about the epistemic and delegative competencies of other participants is not required. Coordination is much simpler, and, in the case of independent competencies, is easily attainable. The external relative weight intervention aspect of transfer delegation creates very substantial epistemic challenges and an accompanying elevated likelihood of failure.

It may be that decentralization requires an assist due to the epistemic difficulties that loom. We considered four possibilities. Systematic independent transfer delegation by whales, in which they delegate substantial voting rights to a panel of delegates who have a high degree of assured independence from the whales, has promise. By carefully selecting the panel to include independent judgments and high epistemic competencies and by appropriately assigning delegation amounts, a whale can construct a big block of votes that comes close to a vote with optimal weights. If this block is substantial enough in light of direct participation by voters outside the block, it may create an approximation to the optimal epistemic environment. If so, direct participation outside of the block is less likely to have significant negative epistemic consequences, creating a happy world in which the potential for conflict between increasing direct participation and achieving epistemic goals is reduced. 

The other three possibilities are based on market approaches. Futarchy uses prediction markets to make discrete decisions, and the associated market aggregation of information may be superior to aggregation through voting mechanisms for some issues. Use of Condorcet AI agents creates the possibility of replicating optimal voting outcomes among a group of such agents, potentially creating a powerful way to implement systematic independent transfer delegation. A contestable control mechanism leverages the token market to create guardrails for DAO governance processes. The epistemic quality of this mechanism depends on the ability of token prices to represent the quality of DAO performance in terms of the DAO objectives, whether or not commercial. The mechanism creates an auction designed to execute the best possible plan going forward in terms of token market capitalization. If governance has gone astray because of epistemic defects, the mechanism provides a path for a reset or reform of the governance system. In that case the potential plans that are the subject of the auction are governance plans. 

Many questions flow from what we have discussed: How close do we need to be to optimal voting weights to ameliorate possible epistemic losses stemming from various kinds of increased participation? Are there methods of moving toward optimal weights that are related to independent transfer delegation by whales but that have more a decentralized nature? Will those methods work without the incentives that stem from substantial token ownership? Can transfer delegation approaches be devised that overcome some of the epistemic weaknesses and difficulties that appear to be present? Would an assist from AI agents help under any of the approaches? Does the partial abstention approach have practical utility? 

Because epistemic performance is central to the survival and flourishing of any governance institution, answers to these questions in the case of DAO governance would be immensely valuable. The epistemic tests we have employed in this paper have helped to frame and sharpen these questions. Although epistemic tests cannot on their own provide solutions guaranteed to work in practice, we believe that researchers and DAO participants should consider them as potentially very useful tools to advance the understanding of decentralized governance.



\newpage
\bibliographystyle{econ_no_doi} 
\bibliography{Econ_and_Game_Theory}

@Article{Bar_Isaac_2020,
  author    = {Heski Bar-Isaac and Joel Shapiro},
  journal   = {Journal of Financial Economics},
  title     = {Blockholder voting},
  year      = {2020},
  issn      = {0304-405X},
  month     = jun,
  number    = {3},
  pages     = {695--717},
  volume    = {136},
  doi       = {10.1016/j.jfineco.2019.11.005},
  publisher = {Elsevier {BV}},
}

@Article{Jensen_1976,
  author    = {Michael C. Jensen and William H. Meckling},
  journal   = {Journal of Financial Economics},
  title     = {Theory of the firm: Managerial behavior, agency costs and ownership structure},
  year      = {1976},
  month     = {oct},
  number    = {4},
  pages     = {305--360},
  volume    = {3},
  doi       = {10.1016/0304-405X(76)90026-X},
  publisher = {Elsevier {BV}},
}

@Article{Nitzan_1982,
  author    = {Shmuel Nitzan and Jacob Paroush},
  journal   = {International Economic Review},
  title     = {Optimal Decision Rules in Uncertain Dichotomous Choice Situations},
  year      = {1982},
  month     = {jun},
  number    = {2},
  pages     = {289},
  volume    = {23},
  doi       = {10.2307/2526438},
  publisher = {{JSTOR}},
}

@Article{Reyes_2017,
  author  = {Carla L. Reyes and Nizan Geslevich Packin and Edwards, Benjamin P.},
  journal = {William and Mary Law Review Online},
  title   = {Distributed Goverance},
  year    = {2017},
  pages   = {1-32},
  volume  = {59},
  url     = {https://scholarship.law.wm.edu/wmlronline/vol59/iss1/1/},
}

@Article{Liu_2023,
  author    = {Xuan Liu},
  journal   = {{SSRN} Electronic Journal},
  title     = {The Illusion of Democracy? An Empirical Study of {DAO} Governance and Voting Behavior},
  year      = {2023},
  note      = {http://dx.doi.org/10.2139/ssrn.4441178},
  comment   = {(May 8, 2023). Available at: https://ssrn.com/abstract=4441178 or http://dx.doi.org/10.2139/ssrn.4441178},
  doi       = {10.2139/ssrn.4441178},
  publisher = {Elsevier {BV}},
  url       = {http://dx.doi.org/10.2139/ssrn.4441178},
}

@Article{Feichtinger_2023,
  author    = {Feichtinger, Rainer and Fritsch, Robin and Vonlanthen, Yann and Wattenhofer, Roger},
  title     = {The Hidden Shortcomings of {(D)AOs} -- An Empirical Study of On-Chain Governance},
  year      = {2023},
  note      = {https://doi.org/10.48550/arxiv.2302.12125},
  copyright = {arXiv.org perpetual, non-exclusive license},
  doi       = {10.48550/arXiv.2302.12125},
  publisher = {arXiv},
  url       = {https://doi.org/10.48550/arxiv.2302.12125},
}

@Book{Condorcet_1785,
  author    = {Marquis de Condorcet},
  publisher = {De l'imprimerie royale (Paris)},
  title     = {Essai sur l'application de l'analyse à la probabilité des décisions rendues à la pluralité des voix},
  year      = {1785},
  url       = {https://gallica.bnf.fr/ark:/12148/bpt6k417181/f4.item},
}

@InCollection{Dietrich_2023,
  author       = {Dietrich, Franz and Spiekermann, Kai},
  booktitle    = {The {Stanford} Encyclopedia of Philosophy},
  publisher    = {Metaphysics Research Lab, Stanford University},
  title        = {{Jury Theorems}},
  year         = {2023},
  edition      = {{S}pring 2023},
  editor       = {Edward N. Zalta and Uri Nodelman},
  note         = {Spring 2023 edition.},
  howpublished = {\url{https://plato.stanford.edu/archives/spr2023/entries/jury-theorems/}},
}

@Misc{HR4763,
  author  = {{House Financial Services Committee}},
  note    = {July 26, 2023.},
  title   = {House {Financial} {Services} {Committee} {Reports} {Digital} {Asset} {Market} {Structure}, {National} {Security} {Legislation} to {Full} {House} for {Consideration}},
  year    = {2023},
  url     = {https://financialservices.house.gov/news/documentsingle.aspx?DocumentID=408940},
  urldate = {2023-09-25},
}

@Article{Buterin_2022_09,
  author  = {Vitalik Buterin},
  title   = {{DAOs} are not corporations: where decentralization in autonomous organizations matters},
  year    = {2022},
  month   = sep,
  note    = {Blog post September 20, 2022.},
  url     = {https://vitalik.eth.limo/general/2022/09/20/daos.html},
  urldate = {2024-03-20},
}

@Article{Hardt_2015,
  author   = {Hardt, Steve and Lopes, Lia C R},
  journal  = {Technical Disclosure Commons: Defensive Publication Series},
  title    = {Google {Votes}: {A} {Liquid} {Democracy} {Experiment} on a {Corporate} {Social} {Network}},
  year     = {2015},
  language = {en},
  url      = {https://www.tdcommons.org/cgi/viewcontent.cgi?article=1092&context=dpubs_series},
}

@Article{Amico_2021,
  author   = {Jeff Amico},
  journal  = {a16z crypto},
  title    = {Open {Sourcing} {Our} {Token} {Delegate} {Program}},
  year     = {2021},
  month    = aug,
  language = {en},
  url      = {https://a16zcrypto.com/posts/article/open-sourcing-our-token-delegate-program/},
  urldate  = {2024-04-06},
}

@Article{Strnad_2024,
  author = {Jeff Strnad},
  title  = {Economic DAO Governance: A Contestable Control Approach},
  year   = {2024},
  note   = {Version 4, December 25, 2024.},
  eprint = {arXiv:2403.16980},
  url    = {https://arxiv.org/abs/2403.16980},
}

@Article{Kahng_2021,
  author    = {Anson Kahng and Simon Mackenzie and Ariel D. Procaccia},
  journal   = {Journal of Artificial Intelligence Research},
  title     = {Liquid Democracy: An Algorithmic Perspective},
  year      = {2021},
  issn      = {1076-9757},
  month     = mar,
  pages     = {1223--1252},
  volume    = {70},
  doi       = {10.1613/jair.1.12261},
  publisher = {AI Access Foundation},
}

@Misc{HR4763pass,
  author  = {{House Financial Services Committee}},
  note    = {May 22, 2024.},
  title   = {House {Passes} {Financial} {Innovation} and {Technology} for the 21st {Century} {Act} with {Overwhelming} {Bipartisan} {Support}},
  year    = {2024},
  url     = {https://financialservices.house.gov/news/documentsingle.aspx?DocumentID=409277},
  urldate = {2024-06-03},
}

@Book{Brennan_2016,
  author    = {Brennan, Jason},
  publisher = {Princeton University Press},
  title     = {Against Democracy},
  year      = {2016},
  isbn      = {9781400882939},
  month     = aug,
  doi       = {10.1515/9781400882939},
}

@Article{Ben_Yashar_2000,
  author    = {Ben-Yashar, Ruth and Paroush, Jacob},
  journal   = {Social Choice and Welfare},
  title     = {A nonasymptotic Condorcet jury theorem},
  year      = {2000},
  issn      = {1432-217X},
  month     = mar,
  number    = {2},
  pages     = {189--199},
  volume    = {17},
  doi       = {10.1007/s003550050014},
  publisher = {Springer Science and Business Media LLC},
}

@Article{Ben_Yashar_2017,
  author    = {Ben-Yashar, Ruth and Nitzan, Shmuel},
  journal   = {Public Choice},
  title     = {Are two better than one? A note},
  year      = {2017},
  issn      = {1573-7101},
  month     = mar,
  number    = {3–4},
  pages     = {323--329},
  volume    = {171},
  doi       = {10.1007/s11127-017-0439-7},
  publisher = {Springer Science and Business Media LLC},
}

@Article{Paroush_1997,
  author    = {Paroush, Jacob},
  journal   = {Social Choice and Welfare},
  title     = {Stay away from fair coins: A Condorcet jury theorem},
  year      = {1997},
  issn      = {1432-217X},
  month     = nov,
  number    = {1},
  pages     = {15--20},
  volume    = {15},
  doi       = {10.1007/s003550050088},
  publisher = {Springer Science and Business Media LLC},
}

@Article{Novak_2024,
  author  = {Graham Novak},
  journal = {X},
  title   = {Introducing Rank-Choice Delegation},
  year    = {2024},
  month   = aug,
  note    = {Posted on X, August 6, 2024.},
  url     = {https://x.com/gnovak_/status/1820784566426018074},
}

@Article{Mooers_2024,
  author        = {Victoria Mooers and Joseph Campbell and Alessandra Casella and Lucas de Lara and Dilip Ravindran},
  title         = {Liquid Democracy. Two Experiments on Delegation in Voting},
  year          = {2024},
  note          = {arXiv:2212.09715v2},
  archiveprefix = {arXiv},
  eprint        = {2212.09715},
  primaryclass  = {econ.GN},
  url           = {https://arxiv.org/abs/2212.09715},
}

@Article{Dietrich_2024,
  author    = {Dietrich, Franz and Spiekermann, Kai},
  journal   = {Economic Theory},
  title     = {Deliberation and the wisdom of crowds},
  year      = {2024},
  issn      = {1432-0479},
  month     = jul,
  number    = {2},
  pages     = {603--655},
  volume    = {79},
  doi       = {10.1007/s00199-024-01595-4},
  publisher = {Springer Science and Business Media LLC},
}

@Article{Hall_2024,
  author = {Andrew Hall and Sho Miyazaki},
  title  = {What Happens When Anyone Can Be Your Representative? {S}tudying the Use of Liquid Democracy for High-Stakes Decisions in Online Platforms},
  year   = {2024},
  month  = oct,
  note   = {October 30, 2024},
  url    = {https://www.gsb.stanford.edu/faculty-research/working-papers/what-happens-when-anyone-can-be-your-representative-studying-use},
}

@Article{Bloembergen_2019,
  author  = {Daan Bloembergen and Davide Grossi and Martin Lackner},
  journal = {The Thirty-Third AAAI Conference on Artificial Intelligence (AAAI-19)},
  title   = {On Rational Delegations in Liquid Democracy},
  year    = {2019},
  pages   = {1796-1803},
}

@InCollection{Escoffier_2019,
  author    = {Escoffier, Bruno and Gilbert, Hugo and Pass-Lanneau, Adèle},
  booktitle = {Algorithmic Game Theory},
  publisher = {Springer International Publishing},
  title     = {The Convergence of Iterative Delegations in Liquid Democracy in a Social Network},
  year      = {2019},
  isbn      = {9783030304737},
  pages     = {284--297},
  doi       = {10.1007/978-3-030-30473-7_19},
  issn      = {1611-3349},
}

@Article{Lever_2023,
  author  = {Annabelle Lever},
  journal = {Danish Yearbook of Philosophy},
  title   = {Democracy: Should We Replace Elections withRandom Selection?},
  year    = {2023},
  pages   = {136–153},
  volume  = {56},
}

@Misc{Butter_2025,
  author  = {Butter},
  note    = {Accessed February 11, 2025.},
  title   = {Butter},
  year    = {2025},
  comment = {Available at https://butterd.notion.site/.},
  url     = {https://butterd.notion.site/},
}

@Article{Hanson_2013,
  author    = {Hanson, Robin},
  journal   = {Journal of Political Philosophy},
  title     = {Shall We Vote on Values, But Bet on Beliefs?},
  year      = {2013},
  issn      = {1467-9760},
  month     = feb,
  number    = {2},
  pages     = {151--178},
  volume    = {21},
  doi       = {10.1111/jopp.12008},
  publisher = {Wiley},
}

@InCollection{Snowberg_2013,
  author    = {Snowberg, Erik and Wolfers, Justin and Zitzewitz, Eric},
  booktitle = {Handbook of Economic Forecasting},
  publisher = {Elsevier},
  title     = {Prediction Markets for Economic Forecasting},
  year      = {2013},
  chapter   = {11},
  editor    = {Graham Elliott, Allan Timmermann},
  isbn      = {9780444536839},
  pages     = {657--687},
  doi       = {10.1016/B978-0-444-53683-9.00011-6},
  issn      = {1574-0706},
}

@Book{Dixon_2024,
  author    = {Chris Dixon},
  publisher = {Random House},
  title     = {Read Write Own: Building the Next Era of the Internet},
  year      = {2024},
}

@InCollection{Airiau_2024,
  author    = {Airiau, Stéphane and Dupuis, Nicholas Kees and Grossi, Davide},
  booktitle = {Algorithmic Game Theory},
  publisher = {Springer Nature Switzerland},
  title     = {Condorcet Markets},
  year      = {2024},
  isbn      = {9783031710339},
  pages     = {501--519},
  doi       = {10.1007/978-3-031-71033-9_28},
  issn      = {1611-3349},
}

@Article{Feddersen_1996,
  author  = {Timothy J. Feddersen and Wolfgang Pesendorfer},
  journal = {American Economic Review},
  title   = {The Swing Voter's Curse},
  year    = {1996},
  number  = {3},
  pages   = {408-424},
  volume  = {86},
}

@Article{Spannocchi_2025,
  author  = {Raphael Spannocchi},
  journal = {X},
  title   = {A Taxonomy of Voter Intent},
  year    = {2025},
  month   = mar,
  note    = {Posted on X, March 3, 2025.},
  url     = {https://x.com/raphbaph/article/1896538021081583882},
}

@Article{Fritsch_2024,
  author    = {Fritsch, Robin and Müller, Marino and Wattenhofer, Roger},
  journal   = {Blockchain: Research and Applications},
  title     = {Analyzing voting power in decentralized governance: Who controls {DAOs}?},
  year      = {2024},
  issn      = {2096-7209},
  month     = sep,
  note      = {Article 100208},
  number    = {3},
  volume    = {5},
  doi       = {10.1016/j.bcra.2024.100208},
  publisher = {Elsevier BV},
}

@Article{Strnad_2025cc,
  author  = {Jeff Strnad},
  journal = {Blockchain: Research and Applications},
  title   = {Economic DAO Governance: A Contestable Control Approach},
  year    = {2025},
  note    = {forthcoming},
}

@InCollection{Khanna_2022,
  author    = {Vikramaditya S. Khanna},
  booktitle = {The Cambridge Handbook of Shareholder Engagement and Voting},
  publisher = {Cambridge University Press},
  title     = {Shareholder Engagement in the {United States}},
  year      = {2022},
  chapter   = {12},
  editor    = {Harpreet Kaur, Chao Xi, Christoph Van der Elst, Anne Lafarre},
  month     = {sep},
  pages     = {239--260},
  doi       = {10.1017/9781108914383.013},
}

\end{document}